\documentclass{jfm}
\usepackage{graphicx}

\usepackage{epstopdf, epsfig}
\usepackage[usenames,dvipsnames]{xcolor}

\usepackage{amsmath}

\usepackage{lastpage}
\usepackage[format=plain,justification=justified,singlelinecheck=false,font={stretch=1.125,small,sf},labelfont=bf,labelsep=space,center]{caption}
\usepackage{amsmath,bm}
\usepackage{fancyhdr}
\usepackage{fnpos}
\usepackage[english]{babel}
\addto{\captionsenglish}{%
  
}
\usepackage{array}
\usepackage{droidsans}
\usepackage{charter}
\usepackage[T1]{fontenc}
\usepackage[usenames,dvipsnames]{xcolor}
\usepackage{setspace}
\usepackage[compact]{titlesec}
\usepackage[hidelinks]{hyperref}

\usepackage{placeins}
\usepackage{epstopdf}
\usepackage{enumerate}
\usepackage{paralist}

\usepackage{subfig}
\usepackage{stackengine}
\usepackage{placeins}

\def\O{\mathcal{O}}
\def\e{\varepsilon}
\def\g{\dot{\gamma} }
\def\Ca{\Ca }
\def\C{\mathcal{C}}

\def\PARAM{\Upsilon}

\def\that{\bm{\hat{\theta}}}
\def\rhat{\bm{\hat{r}}}
\def\shat{\bm{\hat{s}}}
\def\nhat{\bm{\hat{n}}}
\def\xhat{\bm{\hat{x}}}
\def\yhat{\bm{\hat{y}}}
\def\zhat{\bm{\hat{z}}}

\def\x{\bm{x}}
\def\r{\bm{r}}
\def\u{\bm{u}}
\def\F{\bm{F}}
\def\f{\bm{f}}
\def\M{\bm{M}}

\def\I{\mathbf{I}}

\def\ro{a_0}
\def\rcn{a_{n}}
\def\rsn{b_{n}}

\def\rhotwo{\rho^{(2)}}

\def\rcntwo{a^{(2)}_{n}}
\def\rsntwo{b^{(2)}_{n}}

\def\po{\phi^{(0)}}
\def\bkc{\bar{\kappa}\C^{-1}}

\def\roeq{\tilde{a}_0}
\def\rstwoeq{\tilde{b}_{2}}

\def\rctwotwoeq{\tilde{a}^{(2)}_{2}}
\def\rcfourtwoeq{\tilde{a}^{(2)}_{4}}

\def\Ca{\ensuremath \textnormal{Ca}}

\newcommand{\tcb}{\textcolor{black}}

\shorttitle{Swinging and tumbling of multicomponent vesicles in flow}
\shortauthor{P. Gera, D. Salac and S. E. Spagnolie}

\title{Swinging and tumbling of multicomponent vesicles in flow}

\author{Prerna Gera\aff{1},
  David Salac\aff{2}
 \and Saverio E.  Spagnolie\aff{1,3}\corresp{\email{spagnolie@math.wisc.edu}}}

\affiliation{\aff{1}Department of Mathematics, University of Wisconsin-Madison,
Madison, WI 53711, USA
\aff{2}Department of Mechanical Engineering, University of
Buffalo, Buffalo, NY, USA
\aff{3}Dept. Chem. \& Biol. Engineering, University of Wisconsin-Madison,
Madison, WI 53711, USA
}

\graphicspath{{figures/}}
\begin{document}

\maketitle

\begin{abstract}
Biological membranes are host to proteins and molecules which may form domain-like structures resulting in spatially-varying material properties. Vesicles with such heterogeneous membranes can exhibit intricate shapes at equilibrium and rich dynamics when placed into a flow. Under the assumption of small deformations we develop a reduced order model to describe the fluid-structure interaction between a viscous background shear flow and an inextensible membrane \tcb{in two dimensions} with spatially varying bending stiffness and spontaneous curvature. Material property variations of a critical magnitude, relative to the flow rate and internal/external viscosity contrast, can set off a qualitative change in the vesicle dynamics. A membrane of nearly constant bending stiffness or spontaneous curvature undergoes a small amplitude swinging motion (which includes tangential tank-treading), while for large enough material variations the dynamics pass through a regime featuring tumbling and periodic phase-lagging of the membrane material, and ultimately for very large material variation to a rigid body tumbling behavior. Distinct differences are found for even and odd spatial modes of domain distribution. Full numerical simulations are used to probe the theoretical predictions, which appear valid even when studying substantially deformed membranes.
\end{abstract}


\renewcommand{\vec}[1]{\ensuremath\mathbf{#1}}
\newcommand{\nc}{\textcolor{red}{[NC]}}
\newcommand{\iap}[1]{\textcolor{blue}{#1}}
\newcommand{\iapeq}[1]{\ensuremath\textcolor{blue}{#1}}
\newcommand{\red}[1]{\textcolor{red}{#1}}

\section{Introduction}

Biological membranes are often modeled as being homogeneous in composition, a simplification which has resulted in a trove of understanding of their shapes, dynamics in flows, fission, and beyond. But real biological membranes contain a vast array of proteins which can form domains resulting in spatial variations in material properties, leading to changes in vesicle shapes~\citep{Seifert97,hwl11}. Simpler systems of synthetic multicomponent vesicles, whose membranes can be composed of different lipid species, have been used to study the rich patterns and accompanying morphologies which emerge from elastic heterogeneity~\citep{vk03, bhw03}. These findings have been corroborated and expanded upon using both numerical and analytical techniques~\citep{es13,bgn17}, which in turn are of use when attempting to infer membrane properties experimentally~\citep{eds85,bdwj05,tjwb07}.

The analytical study of vesicles with single-component membranes in flow has a long history. \cite{ks82} demonstrated with a two-dimensional elliptical membrane a transition from tank-treading (in which the membrane shape and orientation are fixed but the membrane material slides along the surface) to tumbling (the long axis rotates in a periodic fashion) beyond a critical interior/exterior viscosity contrast. But in general a vesicle is not ellipsoidal and must be determined through a balance of interfacial forces, for instance by describing the shape using a series expansion~\citep{br81,zrs87}. \tcb{Barthes-Biesel~\citep{Barthes-Biesel80,br81,Barthes-Biesel91} also considered the impact of the internal/external viscosity ratio for nearly spherical capsules assuming zero membrane bending stiffness.} More recently, \cite{Misbah06} extended the Keller-Skalak model by including bending rigidity and assuming small deformations, and unearthing a dynamic mode called ``vacillating-breathing''  (the membrane orientation undergoes oscillation around the flow direction while the axis lengths vary in time). In the same year an experimental work by \cite{ks06} identified a ``trembling'' mode (shape deformation oscillations), followed not long after by observation of vesicle ``swinging" (periodic oscillations about a fixed orientation) by \cite{ng07}. \tcb{There appears to be some disagreement in the literature about the precise meaning of these definitions, and whether or not these modes are independent. Some of the above terms are used interchangeably by various authors (a semantic issue also noted by \cite{Misbah12}).}

Phase diagrams for the shapes and dynamics of vesicles in linear flows has been mapped out by numerous authors \citep{dkss09, dks09,zs11,zsdks11,alss14}; see also \cite{Barthes-Biesel16}. The roles of nearby boundaries \citep{zss11}, inertia \citep{sm12}, semi-permeability \citep{qgy21}, enclosed particles \citep{vyvb11}, fluid viscoelasticity \citep{mpwsa16,stkl19}, thermal fluctuations \citep{wjs97,sjw84,mm94,mbf94,Seifert99,flsg08,as16}, and active internal stresses \citep{gl17,yss2021} are among the many additional physical and biological features that have been considered, and a large body of literature is devoted to suspensions of many \tcb{deformable particles such as cells and} vesicles in flows \citep{kss08,vpm09,vrbz11,zsn12,Freund14,kg15,rda19}. \tcb{The membrane viscosity itself, meanwhile, can be accessed using flow patterns in the membrane driven by viscous stresses \citep{hwkg13}}. A more comprehensive review is provided by \cite{alss14}.

The behaviors of multicomponent vesicles in flows, meanwhile, has only just begun to attract attention. Analytical results are scarce but numerical simulations have offered substantial insight. Simulations in a stationary environment have revealed wrinkling and budding deformations \citep{llv12}, and the formation of multicomponent vesicles by adhesion and fusion \citep{zd11}. \cite{stlvl10} studied two-dimensional multicomponent vesicles in a background shear flow, along with the evolution of distinct surface phases, finding highly complex morphologies and dynamics for highly deformed vesicles. The influence of both bending rigidity and spontaneous curvature variation on the equilibrium shape of vesicle has also been investigated \citep{Cox_2015}. Subsequent boundary integral simulations by \cite{lmalvl17} showed a transition from tumbling to tank-treading to ``phase-treading'' of the constituents along the surface upon increasing the shear-rate. Analytic results have also shown that a variation of bending rigidity along a surface can induce migration in tank-treading vesicles~\citep{PIERO2011}.

Synthetic systems have also been fruitful for testing theoretical predictions. Experiments using a two-phase lipid vesicle in such a flow as a simplified model of red blood cell dynamics showed similarly complex features~\citep{afv07,av08,dsv12,vbm13,sg15,tlakhv18}. \cite{gs18} then used simulations to probe a wide array of morphological changes due to spatially varying bending stiffness and line tension between two lipid phases. The phase separation process itself is naturally of great interest, and experiments have been used to study spinodal decomposition and viscous fingering along membrane surfaces~\citep{vk03,lrv09,mo13,shpwlmmhk13}.

In this article we derive analytical predictions for a two-dimensional, multicomponent vesicle in a shear flow under the assumption of small deformations and already-formed domains. Among the fruits of the reduced-order model so produced is a single equation describing the inclination angle dynamics when the distribution of material properties varies in the second spatial mode, the frequency in which they interact most strongly with the extensional part of the background flow. In this most dynamic case, sharp transitions from swinging with tank-treading to tumbling is identified, passing through a transition regime with periodic phase-lagging of the material relative to the vesicle's elongated axis. The method of matched asymptotics is used to produce an approximate solution to the inclination angle equation through this sharp transition, as well as the critical value of the bifurcation parameter signaling the transition from swinging to tumbling which depends on the material property gradient, shear-rate, and internal/external viscosity contrast. The asymptotic predictions are shown to compare favorably to the results of full numerical simulations, even for highly deformed vesicles.

The paper is organized as follows. After presenting the mathematical framework in \S\ref{S: model} to describe the coupling of the fluid flow and elastic membrane stresses at zero Reynolds number (Stokes flow), an expansion is performed around a nearly circular vesicle to reduce the system down to time-dependent shape equations. The classical case of constant membrane material properties is presented in \S\ref{S: constant bending}, in which the results of asymptotic predictions are compared to full numerical simulations. In \S\ref{S: variable bending} attention is turned to the case of interest, that of spatially varying material properties, in which the resulting dynamics are shown to depend strongly on the spectrum of the material properties, and in particular the parity of the number of domains. Concluding remarks are provided in \S\ref{S: conclusion}.

\section{Mathematical model}\label{S: model}
\subsection{Membrane shape and small deformations}

The membrane, or vesicle surface, $S$, is described by a surface parameterization $\r (s,t)$, where $s$ is the arc-length and $t$ is time. The unit tangent and outward-pointing normal vectors on the surface are written as $\shat=\r_s$ and $\nhat = \shat^\perp$. The membrane is assumed area-preserving with area $A$ and inextensible with length $L=2\pi a$ (so that $s\in[0,L)$).

\begin{figure}
\centering
\includegraphics[width=.8\textwidth]{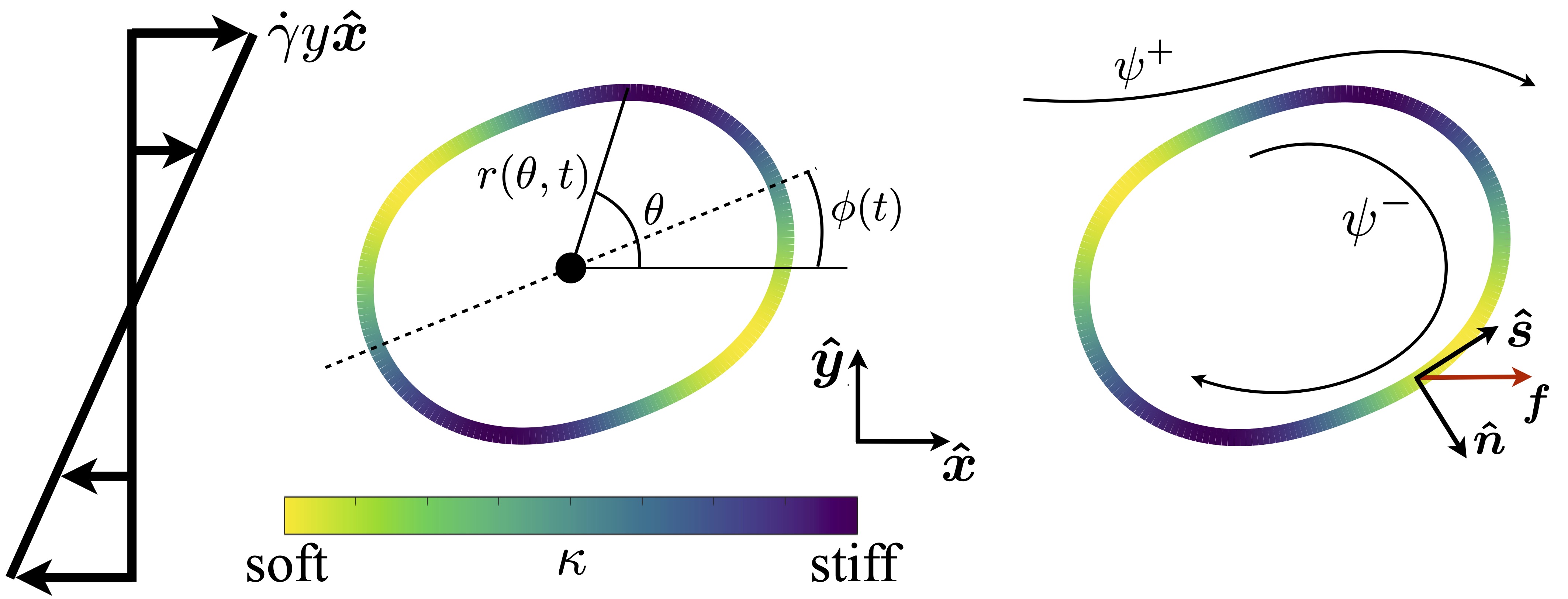}
\caption{(Color online) (Left) Schematic of the two-dimensional inextensible membrane with spatially varying bending stiffness (or spontaneous curvature) in a shear flow, $\g y\xhat$; $\kappa$ denotes the spatial variation in bending stiffness in Eqn.~\eqref{eq: achi}. Softer regions are lighter in color than the darker, stiffer domains. Here the material properties vary in the second spatial mode (e.g. the membrane has two stiffer domains). (Right) The stream-function for the external and internal flows are denoted by $\psi^+$ and $\psi^{-}$, respectively; the viscous traction, $\f$, in Eqn.~\eqref{eq: traction}, instantaneously balances the elastic traction in Eqn.~\eqref{eq:elastictraction}.}
\label{fig: schematic}
\end{figure}

In the event that the membrane area is not far removed from that of a circle of length $L$, it becomes convenient to work in polar coordinates $(r,\theta)$, with unit vectors $\rhat$ and $\that$, and we represent the surface $S$ as $\r (s(\theta,t),t)= r(\theta,t) \rhat(\theta) = a(1+\e \rho(\theta,t)+\e^2\rhotwo(\theta,t) ) \rhat(\theta)$, where $\e$ is a small non-negative constant. For small $\e$ we have $\shat= \that+\e \rho_\theta\rhat  +\O(\e ^2)$ and $\nhat=\rhat-\e\rho_\theta\that+\O(\e ^2)$. A schematic is provided in Fig.~\ref{fig: schematic}. Fourier series representations of the shape functions $\rho$ and $\rhotwo$ are given by
\begin{gather}
\rho(\theta,t)= \sum_{n=0}^{\infty}\rcn (t)\cos(n \theta)+\rsn (t)\sin(n \theta),\label{eqn:rhodef}
\end{gather}
with a similar expression for $\rhotwo(\theta,t)$ with coefficients $\rcntwo(t)$ and $\rsntwo(t)$. The length of the membrane may then be written (suppressing the time-dependence for the sake of presentation), for $\e \ll 1$ as
\begin{gather}
L = \int_0^{2\pi} |\r_\theta|\,d\theta = 2 \pi a\left(1+\e a_0 + \e^2a_0^{(2)}\right)+\frac{\pi a \e^2}{2}\sum_{n=1}^\infty n^2\left(a_n^2+b_n^2\right)+O(\e^3).
\end{gather}
Fixing the membrane length to $2\pi a$ thus requires that $\ro=0$ and
\begin{gather}
a_0^{(2)} = -\sum_{n=1}^\infty \frac{n^2}{4}\left(a_n^2+b_n^2\right),
\end{gather}
and the enclosed area may in that case be written as
\begin{gather}
A=\int^{2\pi}_{0} \frac{r^2}{2}d\theta
=\pi a^2\left(1-\frac{\e^2}{2}\sum_{n=1}^{\infty}\left(n^2-1\right)\left(\rcn ^2+\rsn ^2 \right)\right)+\O(\e^3).\label{eq: A}
\end{gather}
The constant $\e$ may be written in terms of the area enclosed by the membrane if desired as $\e=(2/3)^{1/2}(1-R_A)^{1/2}/Q$, where $R_A = A/(\pi a^2)$ is the ``reduced area'' (equal to unity when the membrane is circular) and
\begin{gather}
Q=\left(\frac13\sum_{n=1}^{\infty}(n^2-1)(\rcn ^2+\rsn ^2)\right)^{1/2}.
\end{gather}
The value of $Q$ must be constant in time if the dynamics are area preserving. The Fourier contributions at mode $n=1$ correspond to translation of the vesicle without shape-change up to $\O(\e^3)$, and hence do not contribute in the expression above.

\subsection{Stokes equations and viscous traction}
The incompressible Stokes equations describing viscous flow both outside ($+$) and inside ($-$) the vesicle are given by
\begin{gather}
\nabla \cdot \bm{\sigma}^\pm=\bm{0},\,\,\, \, \nabla \cdot \u =0,
\end{gather}
where $\u(\x,t)$ is the fluid velocity a point $\x=(x,y)$ at time $t$ and $\bm{\sigma}^\pm = -p^{\pm}\I+\mu^{\pm}(\nabla\u ^{\pm}+\nabla^{T}\u ^{\pm})$ are the Newtonian stress-tensors for each fluid domain, with $p^\pm$ and $\mu^\pm$ the pressures and fluid viscosities external and internal to the membrane. The undisturbed background flow is a linear, horizontal shear flow with shear-rate $\g$, $\u  = \g y \xhat$, with constant pressure $p_\infty$. A no-slip boundary condition is assumed between the fluid and membrane velocities on both sides of the membrane (there is no relative slipping between the inner and outer membrane surfaces). The local viscous tractions, $\f ^\pm = \pm \nhat \cdot \bm{\sigma}^\pm$, acting on the membrane from the exterior and interior interfaces result in the combined local viscous traction
\begin{gather}\label{eq: traction}
\f  = \nhat\cdot[\bm{\sigma}]_S=-[p]_S\nhat+\nhat \cdot \left[\mu (\nabla\u +\nabla^{T}\u )\right]_S,
\end{gather}
with $[f]_S=\left(f^{+}-f^{-}\right)|_S$ defined to be the jump in $f$ across the boundary $S$.

The continuity equation in the bulk fluid is immediately satisfied with the introduction of a stream-function, $\psi$, defined such that $\u = \nabla^\perp \psi = \psi_y \xhat-\psi_x \yhat$. The Stokes equations then reduce to biharmonic equations interior and exterior to the membrane,
\begin{gather}
\nabla^4 \psi^\pm = 0,\label{Eq:biharmonicStream}
\end{gather}
with $\psi^+ \to \g y^2/2$ as $|\x|\to \infty$, the background shear flow. The general form of the $\theta$-periodic solution to the biharmonic equation is given in \S\ref{AppendixA}. Continuity of velocity across the membrane boundary demands that
\begin{gather}
[\nabla \psi]_S=\bm{0}, \label{Eqn:continuity}
\end{gather}
and surface inextensibility along the membrane demands that
\begin{gather}\label{Eqn: inextensibility}
\nabla_s \cdot \u|_S= \shat\left(\shat\cdot\nabla \right)\cdot \u|_S=\bm{0}
\end{gather}
where $\nabla_s$ is the surface del operator.

\subsection{Force and moment balance}
The membrane is modeled as a thin linearly elastic shell. The bending moment is approximated by $\M=B(s)(H-\tilde{H}(s)) \xhat \times \yhat$, where $B(s)$ and $\tilde{H}(s)$ are the spatially-varying bending stiffness and spontaneous curvature, and $H = \shat\, \cdot \partial_s\nhat$ is the mean curvature. Force and moment balance along the membrane surface at arc-length $s$ are given by $d\M/ds+\shat \times\F =\bm{0}$ and $\f _{elastic}+\f =\bm{0}$, where $\M$ is a thickness-averaged first moment of the elastic stress with units of force, $\f _{elastic}$ is the elastic force per area of the membrane on a surrounding medium, with $\f _{elastic}=d\F /ds$, and $\f $ is the viscous traction acting on the membrane, Eqn.~\eqref{eq: traction}. Defining the tangential component of $\F $ as $T(s)$ (the tension \tcb{per unit length}), we then have
\begin{gather}
\F (s) = T(s)\shat -\shat \times \left(\frac{d}{ds}\left(B(s)(H-\tilde{H}(s))\right)\zhat\right)= T(s)\shat +\nhat \left(B(H-\tilde{H})\right)_s,
\end{gather}
and the elastic force acting on the surrounding medium is then given by
\begin{gather}\label{eq:elastictraction}
\f _{elastic} = \left( - T H +\left(B(H-\tilde{H})\right)_{ss} \right)\nhat + \left(T_s+H\left(B(H-\tilde{H})\right)_s \right) \shat,
\end{gather}
where subscripts indicate partial derivatives. A different expression for the traction appears in the literature starting with the Helfrich free energy which amounts to adding a term $B(s)(H-\tilde{H}(s))^2/2$ to $T$ above \citep{gg17}; see also the review articles by \cite{Powers10} and \cite{Deserno15}. In either case $T$ plays the mathematical role of a Lagrange multiplier which enforces membrane inextensibility.

\subsection{Nondimensionalization}\label{sec: nondim}
\tcb{
A competition of viscous and elastic effects emerges when the stresses associated with the flow, the material property variations, and the shape deformations are all on the same scale. In order to see this more clearly, the system is made dimensionless by scaling lengths by $a$, velocities by $a \g$, forces by $\mu^{+} a^2 \g$, stresses by $\mu^{+} \g$, and energies by $\mu^{+} a^3 \g$, while time is scaled upon $\e/\g$. The remaining dimensionless scalar parameters governing the system are
\begin{gather}
R_A=\frac{A}{\pi a^2},\,\,\,\,  \lambda =\frac{\mu^{-}}{\mu^{+}}, \,\,\,\, \tilde{H}_0  = a \langle \tilde{H}(s)\rangle, \,\,\,\,  \Ca =\displaystyle\frac{\mu^{+} a^3 \g}{\langle B(s)\rangle}, \,\,\,\, \C=\frac{\Ca}{ \e},
\end{gather}
where $R_A$ is the reduced area, $\lambda$ is the inner/outer viscosity ratio, $\tilde{H}_0$ is the mean spontaneous curvature (with $\langle \cdot \rangle$ an average over the membrane perimeter), $\Ca$ is the bending capillary number, and $\C$ is a parameter which is $O(1)$ as $\e\to 0$. In addition to these scalar parameters, and with variations away from their mean values assumed to be small, we have the dimensionless distributions of the bending stiffness and spontaneous curvature along the membrane surface,
\begin{gather}
\frac{B(s(\theta),t)}{\langle B(s)\rangle} =1+\e\kappa(\theta,t),\,\,\,\,
a\, \tilde{H}(s(\theta),t)=\tilde{H}_0 +\e \zeta(\theta,t),\label{eq: achi}
\end{gather}
respectively.} Given the periodicity of the system in $\theta$ we also define $\kappa(\theta,t) = \sum_{n=1}^{\infty} c_{n}(t)\cos(n\theta)+d_{n}(t)\sin(n\theta)$, and $\zeta(\theta,t)$ similarly with coefficients $e_{n}(t)$ and $f_{n}(t)$. Henceforth all variables are understood to be dimensionless.

For a membrane of length $2\pi a\approx 120 \mu$m, and bending rigidity $B\approx 20 k_b T$ as measured for a vesicle composed of DOPC lipids \citep{dngksm16,frgvd20}, and using the viscosity of water, $\mu^{+} \approx 10^{-3}$\,Pa\,s, the bending capillary number $\Ca$ is roughly $100\,\g * [1\, \text{second}]$ (e.g. if $\dot{\gamma}=10^{-1}$s$^{-1}$ then $\Ca\approx 10$). The experimental work of \cite{bdwj05}, where the bending rigidity ratio is approximately $1.25$, corresponds here to $\|\e \kappa\|_\infty \approx 0.1$. The Capillary number is highly sensitive to the size; for instance using a length more appropriate to modeling a red blood cell, $2\pi a\approx 20 \mu$m, and with $B\approx 50 k_b T$ \citep{Evans83}, then $\Ca\approx \g/4$. We proceed with the standard abuse of notation, understanding that all variables are now dimensionless. The dimensionless background flow, for instance, is given by $\u = y \xhat$, and the dimensionless membrane perimeter is $L=2\pi$.

\subsection{Membrane shape dynamics}
In order to compute the dynamics of the membrane shape the traction balance is carried out order by order in $\e$, included as \S\ref{AppendixB}, and the tension so found is used instantaneously to solve for the stream-function, included as \S\ref{AppendixC}. To summarize the results, expansions are written for the pressure, $p=p^{(0)}+\e p^{(1)}+...$, tension $T=T^{(0)}+\e T^{(1)}+...$, and velocity $\u= u_n \nhat+ u_s \shat=\left(u_n^{(1)}+\e u_n^{(2)}+...\right)\nhat+\left(u_s^{(1)}+\e u_s^{(2)}+...\right)\shat$. The normal and tangential components of the velocity are given at leading order by
\begin{gather}
u_n\Big|_{S}=\frac{1}{(1+\lambda)}\sin(2 \theta)- 2 \sum_{n=2}^{\infty} n\left(A_n \cos(n\theta)+B_n \sin(n\theta) \right)+\O(\e),\label{eqn:un}\\
u_s\Big|_{S}=-\frac{1}{2}+\frac{1}{2(1+\lambda)}\cos(2 \theta)-2 \sum_{n=2}^{\infty}\left( B_n \cos(n\theta)- A_n \sin(n\theta) \right)+\O(\e),\label{eqn:us}
\end{gather}
where
\begin{gather}\label{eqn:JnKn}
A_n =\frac{\C^{-1} \left[\alpha_n(t) a_n+(1-\tilde{H}_0 )c_n-e_n\right]}{4(1+\lambda)},\,\,\, B_n =\frac{\C^{-1} \left[\alpha_n(t) b_n+(1-\tilde{H}_0 )d_n-f_n\right]}{4(1+\lambda)}.
\end{gather}
Here we have used the Fourier coefficients for the variations in shape given by $a_n, b_n$, in bending stiffness by $c_n, d_n$, and in spontaneous curvature by $e_n, f_n$, and that $\C=\Ca/\e=O(1)$ as $\e \to 0$, and have defined
\begin{gather}\label{eq:alphan}
\alpha_n(t) =  \C\, P_0(t)+n^2-1.
\end{gather}
The function $P_0(t)$ is the leading-order mean pressure jump across the membrane, or equivalently the scaled mean tension, (the two are bound together by an elastic analogue of the Young-Laplace law) and is given by
\begin{gather}
P_0(t)=\frac{14b_2-\C^{-1}\displaystyle\sum _{n=2}^{\infty}n(2n^2-1)C_n}{\displaystyle\sum _{n=2}^{\infty}n(2n^2-1)(\rcn ^2+\rsn ^2)},\label{eqn:pressure}
\end{gather}
where
\begin{gather}\label{eqcn}
C_n = (n^2-1)(\rcn ^2+\rsn ^2) +(1-\tilde{H}_0 )\left(\rcn c_{n} + \rsn d_{n}\right)-\left(\rcn e_{n} + \rsn f_{n}\right).
\end{gather}
Note that $n=2$ terms are present inside the summations in Eqns.~\eqref{eqn:un}, \eqref{eqn:us}.

Finally, the dynamics of the membrane shape are found using the normal component of the velocity field along the surface. As derived in \S\ref{AppendixD} the shape functions satisfy
\begin{gather}\label{eq:rhotun}
\rho_t = u_n^{(1)}\Big|_S = \psi^{(1)}_\theta\Big|_{r=1},\\
\rhotwo_t = u_n^{(2)}\Big|_S = \psi_\theta^{(2)}+\rho\left( \psi_{r\theta}^{(1)}- \psi_\theta^{(1)}\right)+\rho_\theta \psi_r^{(1)}\Big|_{r=1},
\end{gather}
with no ambiguity about the stream-function (internal or external) owing to the continuity of velocity, Eqn.~\eqref{Eqn:continuity}. The end result is that the Fourier modes describing the membrane shape at first-order in $\e$ evolve according to
\begin{gather}
\frac{d a_n}{dt} = \frac{n\C^{-1}}{2 (1+\lambda)} \left(-\alpha_n(t) a_n +(\tilde{H}_0 -1)c_n+e_n \right)\label{eqn:angen},\\
\frac{d b_n}{dt} =\frac{\delta_{n2}}{1+\lambda}+ \frac{n\C^{-1}}{2 (1+\lambda)}\left(-\alpha_n(t) b_n +(\tilde{H}_0 -1)d_n+f_n \right)\label{eqn:bngen},
\end{gather}
where $\alpha_n(t)$ is given in Eqn.~\eqref{eq:alphan}, and $\delta_{n2}$ is unity when $n=2$ and is zero otherwise. In addition, $a_1(t) = a_1(0)$ and $b_1(t) = b_1(0)$, which represents that the system is insensitive to translations of the membrane in either direction at first order in $\e$ (the body translates along with the background flow with any vertical perturbation but the shape dynamics are unchanged). The $n=2$ mode is special since this corresponds to elongation along the principal direction of the background shear flow, at an angle $\pi/4$ relative to the $x$-axis.

Since $P_0(t)$ depends on the membrane shape, the expressions above are immediately nonlinear, even when only considering the leading order shape dynamics in small $\e$. If the mean spontaneous curvature is unity ($\tilde{H}_0 =1$) the membrane remains close enough to its preferred state at first order in $\e$ so that no additional forces are induced by bending, and only spontaneous curvature variations affect the shape dynamics. For any other mean spontaneous curvature ($\tilde{H}_0  \neq 1$) however, the effects of spontaneous curvature are mathematically indistinguishable from bending stiffness at leading order via \eqref{eqn:angen}-\eqref{eqn:bngen}. For the remainder of the paper we will assume zero spontaneous curvature ($e_n=f_n=0$ for all $n$, and $\tilde{H}_0 =0$), but all of the results to come can be viewed as owing to variations to spontaneous curvature rather than bending stiffness, or any combination thereof.

\section{Dynamics of a membrane with uniform material properties}\label{S: constant bending}

We begin by studying the dynamics of a membrane with uniform bending stiffness ($c_n =d_n=0$ for all $n$, and zero spontaneous curvature). In the steady (moving) state, since $\rcn$ and $\rsn$ are constant in time, the pressure jump $P_0(t)$ in Eqn.~\eqref{eqn:pressure} is also constant in time. The dynamics in \eqref{eqn:angen}-\eqref{eqn:bngen} then reveal that all Fourier components vanish exponentially fast with the exception of $b_2$, leaving the steady shape function $\rho(\theta,t) = \rstwoeq \sin(2\theta)$, with $\rstwoeq$ easily determined using area conservation alone:
\begin{gather}
\e \rstwoeq =\e Q = \left(2/3\right)^{1/2}(1-R_A)^{1/2}.\label{eqn:rstwoeqRA}
\end{gather}
Here $R_A$ is the reduced area, having referenced Eqn.~\eqref{eq: A} when only $b_2$ is non-zero.

In particular, a membrane with an initial shape of the form $\rho(\theta,0)=b\sin(2\theta)$ is instantly in a steady state for any $b$ at first order in $\e$. This corresponds to a tilt angle of $\pi/4$ between the vesicle's elongated axis and the direction of flow. Although the shape is stationary, material is still moving along the tangential direction in a so-called tank-treading motion. In this configuration, the steady-state pressure jump is given by $P_0 = -3\C^{-1}+\rstwoeq^{-1}$.

Since the bending stiffness is uniform we are able to examine the steady shape and orientation to higher order in $\e$. Assuming that the membrane shape has already relaxed to the point that $u_n^{(1)}=0$, and hence $\rho_t=0$ from Eqn.~\eqref{eq:rhotun}, a straight-forward continuation of the regular asymptotic expansion yields equations describing the fluid flow at second order resulting in the normal velocity on the membrane surface
\begin{multline}
u_n^{(2)}= \rstwoeq \cos(2\theta)-\frac{3\rstwoeq^2}{2(1+\lambda)}\left(7\C^{-1}+\rstwoeq^{-1}  \right) \cos(4 \theta)\\
 -\frac{1}{2(1+\lambda)}\sum_{n=2}^{\infty}
n\left((n^2-4)\C^{-1}+ \rstwoeq^{-1}\right)\left( \rcntwo \cos(n\theta)+\rsntwo \sin(n\theta)\right)\label{eqn:un2}.
\end{multline}
(Note that there are $\cos(2\theta)$ and $\cos(4\theta)$ terms inside the infinite sum.) The steady state at second-order is reached once $u_n^{(2)}=0$,
\begin{gather}
\lim_{t\to \infty}\rhotwo(\theta,t)=\roeq^{(2)}+\rctwotwoeq\cos(2\theta)+\rcfourtwoeq\cos(4\theta), \label{eqn:rhotwoeq}
\end{gather}
where
\begin{gather}
\roeq^{(2)}=-(\rstwoeq)^{2},\,\,\,\, \rctwotwoeq=(1+\lambda)(\rstwoeq)^{2},\\
\rcfourtwoeq=-\frac{3(\rstwoeq)^{2}}{4}\left([1+7\rstwoeq\C^{-1}]/[1+12\rstwoeq\C^{-1}] \right),
\end{gather}
with $\rstwoeq$ given in Eqn.~\eqref{eqn:rstwoeqRA}. \tcb{Although we assumed above that $\C = O(1)$ as $\e \to 0$, the limit of infinite Capillary number matches the results of \cite{zrs87} who assumed zero bending stiffness. That the zero-bending stiffness limit is recovered as $\C \to \infty$ likely identifies this as a regular limit and not a singular one, though a more general analysis for arbitrary $\C$ would be needed to make this result rigorous.}

\begin{figure}
\centering
\includegraphics[width=\textwidth]{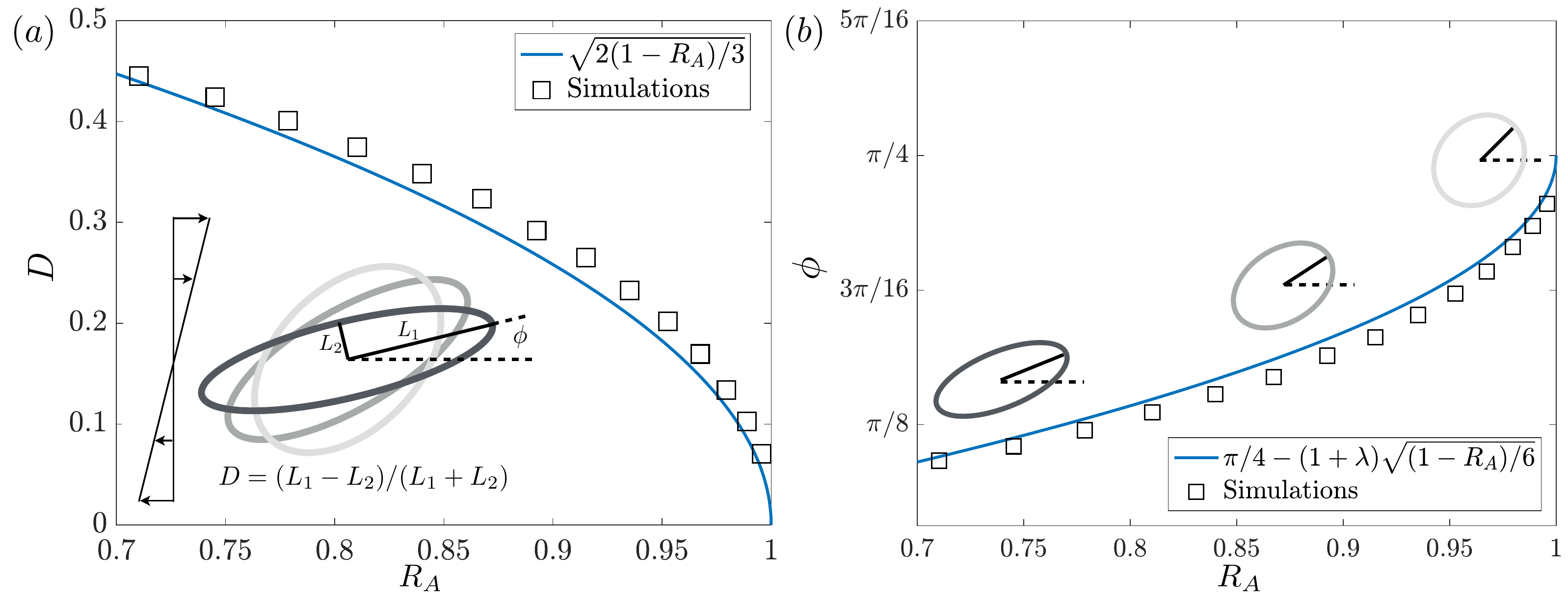}
\caption{The steady-state deformation parameter (a) and inclination angle (b) in the case of constant bending stiffness with viscosity ratio $\lambda=1$ and $\e$ varying from $0$ to $0.15$. Simulations (symbols) and analysis (lines) are in close agreement even for highly deformed membranes.}
\label{fig: constant_bending_comparison}
\end{figure}

\subsection{Steady-State Deformation and Inclination Angle}
The deformation parameter and orientation angle are two common metrics used to characterize the dynamics of a membrane in flow. The Taylor deformation parameter is defined as $D=(L_1-L_2)/(L_1+L_2)$, where $2L_1$ and $2L_2$ are the major and minor axis lengths of an ellipse which shares the same inertia tensor, derived in \S\ref{AppendixE}, resulting as $\e\to 0$ in the representation
\begin{gather}
D(t)= \e  \sqrt{a_{2}^2 + b_{2}^2} +\e ^2 \left(a^{(2)}_{2}a_{2} +b^{(2)}_{2}b_{2}  \right)/\sqrt{a_{2}^2 + b_{2}^2}+\O(\e^3).\label{eqn:TaylorDeform}
\end{gather}
For the case of uniform bending stiffness in the tank-treading steady state,
\begin{gather}
D= \e \rstwoeq +\O(\e^3)=\sqrt{2(1-R_A)/3}+\O(\e^3),\label{eqn:TaylorDeformConstB}
\end{gather}
which is notably independent of any other physics in the problem. The eigenvectors of the inertia tensor, meanwhile, are used to define an inclination angle, $\phi$, the angle between the elongated axis of the membrane and the direction of flow, which has representation (see  \S\ref{AppendixE}):
\begin{gather}
\phi(t)=\arctan{\left( \frac{-a_{2} +\sqrt{a_{2}^2 + b_{2}^2}  }{b_{2}} \right) } +\e \frac{\left(b^{(2)}_{2}a_{2} -a^{(2)}_{2}b_{2} \right)}{2 \left(a_{2}^2 + b_{2}^2  \right)}+\O(\e^2). \label{eqn:incAngle}
\end{gather}
In the case of uniform bending stiffness, in the steady state we find the angle
\begin{gather}\label{phieq}
\phi = \frac{\pi}{4}- \frac{\e(1+\lambda)}{2}\rstwoeq +\O(\e^2)
=\frac{\pi}{4}-(1+\lambda)\sqrt{(1-R_A)/6}+\O(\e^2).
\end{gather}
The predictions above are plotted in Fig.~\ref{fig: constant_bending_comparison} as lines for a range of reduced areas $R_A$.

For nearly circular membranes, the inclination angle approaches $\pi/4$. When fluid is removed from the interior of the membrane the inclination angle decreases and the membrane tilts forward towards the direction of flow. An increase in the viscosity ratio $\lambda=\mu^{-}/\mu^+$ further tilts the membrane down towards the direction of flow. From Eqn.~\eqref{phieq} the critical value of the viscosity ratio for which the steady inclination angle is equal to zero is given by $\pi\sqrt{3}/[2\sqrt{2}\left(1-R_A\right)^{1/2}]-1+\O(\sqrt{1-R_A})$ as $R_A \to 1$. Beyond this critical viscosity ratio the membrane shape is no longer fixed in space and instead undergoes periodic tumbling. The result is independent of the Capillary number, so the same result has been observed in previous work that assumes zero bending rigidity \citep{zrs87}. The result also qualitatively matches the dynamics of a membrane in three-dimensions studied by \cite{vg07}, where the inclination angle was also found to be independent of bending rigidity in the small-deformation regime.

To assess the validity of the asymptotic approximations derived above we solve the complete fluid-structure interaction problem numerically. The incompressible Navier-Stokes equations (which limit to the Stokes equations in Eqn.~\eqref{Eq:biharmonicStream} as the Reynolds number tends to zero) are solved at Reynolds number $10^{-3}$ on a regular grid using a projection method~\citep{ks15} and the vesicle is represented using a semi-implicit level set scheme ~\citep{of02}. A Generalized Minimal Residual algorithm (GMRES) with algebraic multigrid as provided by PETSc \citep{petsc-user-ref,petsc-web-page,petsc-efficient} is used for the level set solver. Derivatives of the level sets are also tracked, in a so-called ``jet''-scheme, to improve the accuracy of interpolants needed to communicate information from the membrane to the fluid and vice-versa~\citep{nrs10,srn12}. More details on the numerical methods used and a convergence study for the code are available in the literature~\citep{vks15,gs18_2}.

Figure~\ref{fig: constant_bending_comparison} includes the results of the full simulations (symbols). The steady-state deformation parameter and inclination angle both show excellent agreement with the numerical simulations (and the predicted order of accuracy as $\e \to 0$, not shown), providing fortuitous accuracy even for substantial membrane deformations where the asymptotic approximations are not immediately expected to hold. The slight overestimate of the deformation parameter for general $\e$ is accompanied by a slight underestimate of the inclination angle, owing to the higher velocities sampled by a more elongated vesicle. In general the transition between tank-treading and tumbling can depend weakly on the bending capillary number, which the above analysis suggests enters at the next order in $\e$~\citep{Lebedev2007,ng10,zs11}.

\section{Dynamics of a membrane with variable material properties}\label{S: variable bending}
If the membrane composition is not uniform, the advection of material around the surface can contribute substantially to the membrane dynamics. Again owing to the mathematically similar contributions of bending stiffness and spontaneous curvature variation we focus our attention on bending stiffness variations. Since the bending stiffness and its spatial variation, $\kappa$, are assumed to be material quantities, they evolve in time according to a surface advection equation which is coupled to the shape equations, introducing a serious analytical challenge. For the sake of tractability, however, we assume that mode-mixing is small and treat $\kappa$ as simply advecting by the mean tangential velocity $-1/2$ in Eqn.~\eqref{eqn:us}. We will see that this approximation leads to predictions that match very well with the results of full numerical simulations. With the bending stiffness variation confined to a single mode $M$ with amplitude $\bar{\kappa}$ (assumed positive), we thus write
\begin{gather}
\kappa(\theta,t)=\bar{\kappa} \cos\left(M(\theta+\e t/2)\right)=  c_{M}(t)\cos(M\theta)+d_{M}(t)\sin(M\theta),\label{eqn:prescribedBending}
\end{gather}
where $c_M(t)=\bar{\kappa}\cos \left(\e M t/2\right)$ and $d_M(t)=-\bar{\kappa}\sin \left(\e Mt/2\right)$. The shape equations are still those in Eqns.~\eqref{eqn:angen}-\eqref{eqn:bngen}, with $c_m$ and $d_m$ also now appearing in Eqn.~\eqref{eqn:pressure}. The cases $M=2$ and $M \neq 2$ are of distinctly different character, and we now proceed to consider them independently.

\subsection{A bifurcation in the dynamic case, $M= 2$:} \label{subsec: bif}
\begin{figure}
\centering
\includegraphics[width=.9\textwidth]{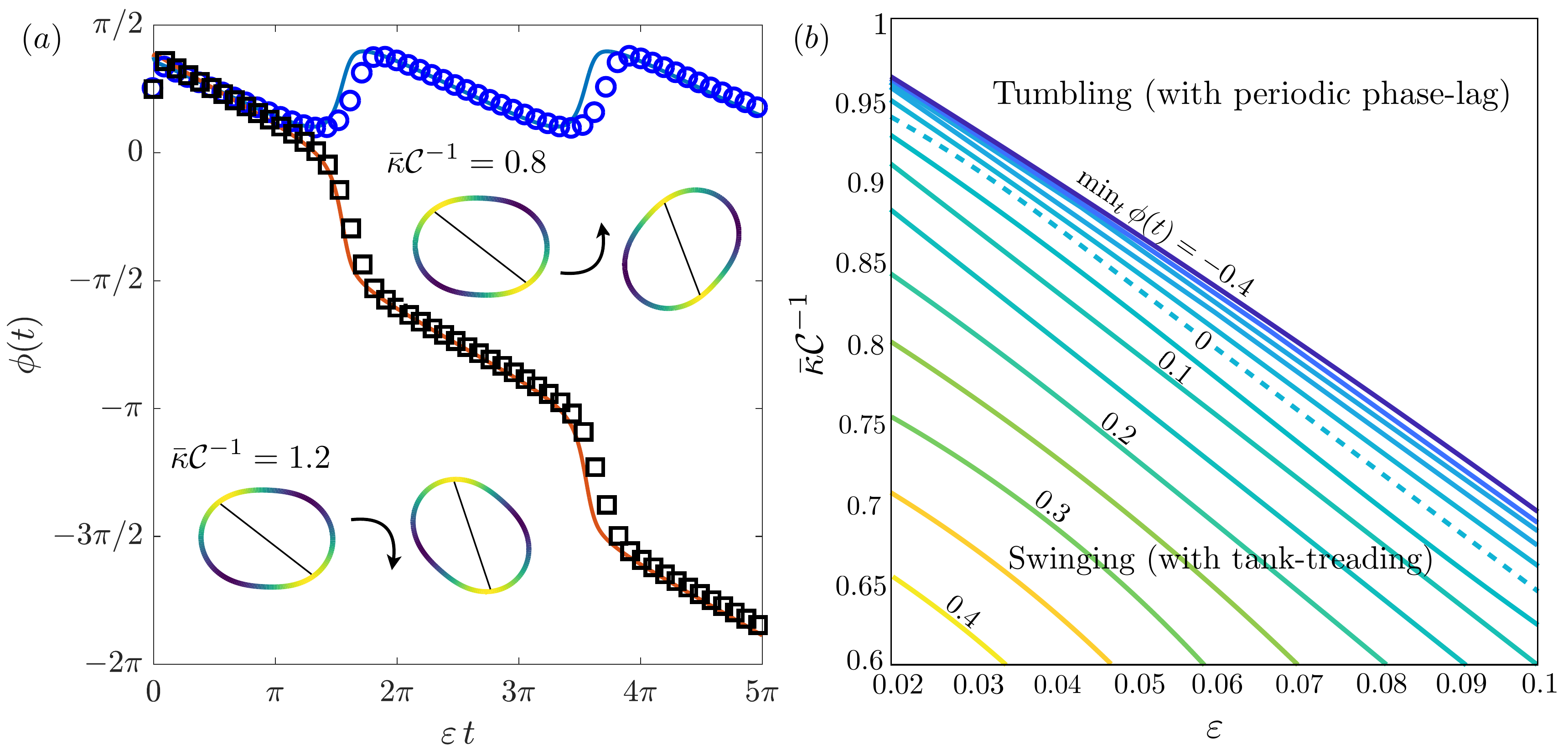}
\caption{(Color online) (a) The inclination angle as a function of time from simulations (symbols) and theory (lines), with $\e=10^{-2}$, $Q=3$, or $R_A=0.998$ and $\lambda=1$. Swinging is observed for $\bkc =0.8$, tumbling for $\kappa=1.2$. Both include additional tank-treading motion, indicated by lines in the snapshots - see also Fig.~\ref{fig: swinging} and Supplemental Movies M1-M3. (b) Contours of the minimum inclination angle during periodic orbits, $\min_t \phi(t)$, computed using the full numerical simulations. \tcb{Note that $\bkc=\e\bar{\kappa}\Ca^{-1}$ so that the vertical axis also depends on $\e$.} }
\label{fig:swinging_to_tumbling}
\end{figure}

\begin{figure}
\includegraphics[width=\textwidth]{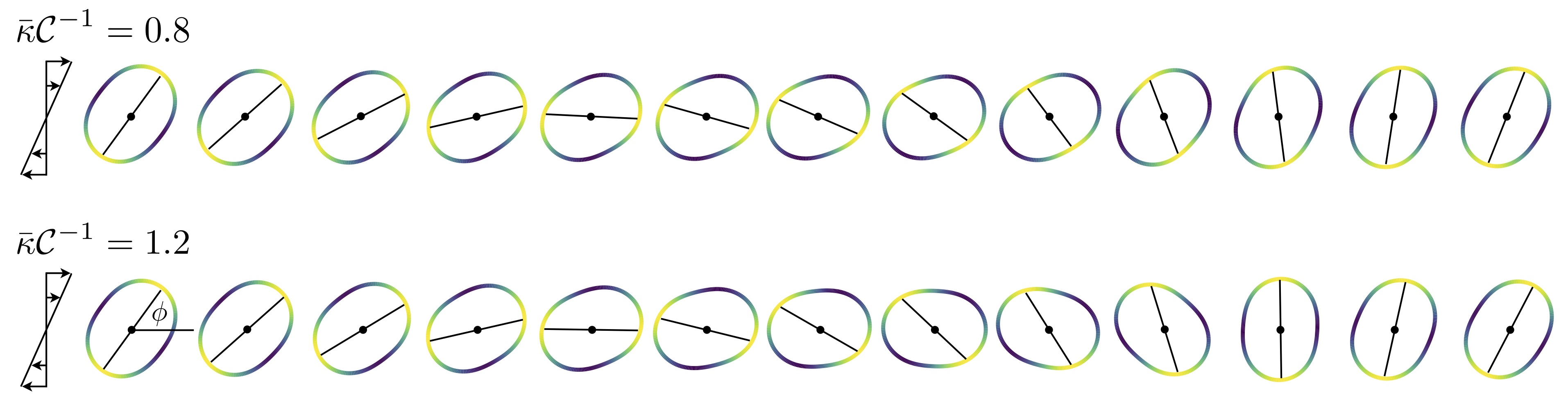}
\caption{(Color online) Snapshots of the dynamics associated with Fig.~\ref{fig:swinging_to_tumbling}(a): swinging with tank-treading ($\bkc=0.8$, top), and tumbling with phase-lagging ($\bkc=1.2$, bottom) with variable bending stiffness in the $M=2$ mode. A line in each snapshot connects the two softer regions, which are lighter in color than the darker, stiffer regions. The vesicle elongates so that the softer regions tend to sit in large curvature regions, while the principal direction of the background flow stretches the vesicle towards inclination angle $\pi/4$. See also Supplemental Movies M1-M3.}
\label{fig: swinging}
\end{figure}

The situation in which the bending stiffness variation is present in the second spatial mode (i.e. two stiff domains, as in Fig.~\ref{fig: schematic}) is a special case, as this is where the distribution of material properties most strongly interacts with the elongating deformation induced by the flow. From Eqns.~\eqref{eqn:angen}-\eqref{eqn:bngen} the shape dynamics in the second mode evolve according to
\begin{gather}
\frac{d a_2}{dt}=\frac{\C^{-1}}{1+\lambda}\left(-\alpha_2(t)a_{2}-\bar{\kappa}\cos\left(\e t\right)\right),\\
\frac{d b_2}{dt}=\frac{1}{1+\lambda}+\frac{\C^{-1}}{1+\lambda}\left(\alpha_2(t)b_2+\bar{\kappa}\sin\left(\e t\right)\right),
\end{gather}
where $\alpha_2(t)=3+\C P_0(t)$. Inserting $P_0(t)$, or equivalently solving for $\alpha_2(t)$ so that $d(a_2^2+b_2^2)/dt=d(Q^2)/dt=0$, the above simplify to
\begin{gather}\label{eqndsig}
\frac{d a_2}{dt}=-\eta(a_2,b_2,t)b_2,\,\,\, \frac{d b_2}{dt}=\eta(a_2,b_2,t)a_2,
\end{gather}
where
\begin{gather}
\eta(a_2,b_2,t) =(1+\lambda)^{-1}Q^{-2}\left[\left(1+\bkc \sin\left(\e t\right)\right)a_2+\bkc \cos(\e t)b_2\right].
\end{gather}
Recall that $Q$ is a constant which is set at $t=0$; if the initial shape deformation resides only in the second Fourier mode, for instance, then $Q=(a_2(0)^2+b_2(0)^2)^{1/2}$. At first order in $\e$ there is no change in the deformation parameter: $D(t) = (2/3)^{1/2}(1-R_A)^{1/2}+\O(\e^2)$, so the observed \textit{shape} does not exhibit large variations in time. The inclination angle, however, reveals something striking. Writing $(a_2,b_2)=Q(\cos 2\po,\sin 2\po)$ (the inclination angle is given by $\phi = \po+O(\e)$) and inserting into Eqn.~\eqref{eqndsig}, an equation for $\po$ arises:
\begin{gather}
\po_t = \beta\left(\cos(2\po)+\bkc \sin(2\po+\e t)\right),\label{eqn:phitO1M=2}
\end{gather}
with $\beta=(1+\lambda)^{-1}Q^{-1}$. This equation is more constructively analyzed by defining the slower timescale $\tau=\e t$, so that
\begin{gather}\label{po2}
\e\po_\tau = \beta\left(\cos(2\po)+\bkc \sin(2\po+\tau)\right).
\end{gather}
Numerical solutions of Eqn.~\eqref{po2} for $\bkc=0.8$ and $\bkc=1.2$ are shown as lines in Fig.~\ref{fig:swinging_to_tumbling}(a), for $\e=10^{-2}$, $\lambda=1$, and $Q=3$. The dynamics alternate between a slow linear drift of $\po(\tau)$ where $\po_\tau=O(1)$ and a rapid departure when $\po$ is near zero. Figure~\ref{fig: swinging}, along with Supplemental Movies M1-M3, show the complex dynamics associated with these plots. When $\bkc =0.8$ the elongated axis swings back and forth relative to the direction of flow; when $\bkc=1.2$ the shape slowly nears a zero inclination angle, then undergoes a rapid tumble. Also shown in Fig.~\ref{fig:swinging_to_tumbling}(a) as symbols are the results found using the full numerical simulations, as in \S\ref{S: constant bending}, showing close agreement with the solutions generated by Eqn.~\eqref{po2}.

When $\bkc$ is small the bending stiffness variation only introduces a periodic perturbation of the constant bending stiffness dynamics. Writing $\po=\pi/4+\bar{\kappa}\C^{-1} \xi(\tau)$ as $\bar{\kappa}\C^{-1}\to 0$ and linearizing Eqn.~\eqref{po2}, we arrive at the periodic solution
\begin{gather}\label{posol}
\po(\tau) = \frac{\pi}{4}+\frac{\bar{\kappa}}{2\C}\cos(\tau)+\O\left(\left(\bar{\kappa}\C^{-1}\right)^2,\e\right) \,\,\,\,\, \mbox{as}\,\,\,\,\bar{\kappa}\C^{-1} \to 0,
\end{gather}
whose period, $\Delta\tau=2\pi$ (or $\Delta t = 2\pi/\e$), is twice that of the material's tangential motion along the surface (since the mean surface tangential velocity is $-\e/2$), owing naturally to the number (two) of stiffer domains. Material tank-treads tangentially along the membrane while the shape swings back and forth. The rapid slipping in Fig.~\ref{fig:swinging_to_tumbling}(a) for the smaller $\bkc$ value occurs when the stiffer material passes quickly over the region of highest curvature.

For very large values of $\bar{\kappa}\C^{-1}$ the dynamics limit to a pure (rigid body) tumbling motion. Assuming a regular perturbation expansion in small $\C/\bar{\kappa}$, the inclination angle has the asymptotic behavior
\begin{gather}
\po(\tau) = \frac{\pi}{2}-\frac{\tau}{2}-\frac{\C}{4\bar{\kappa}}\left(2\cos (\tau)-\e(1+\lambda) Q\right)+\O\left(\left(\C/\bar{\kappa}\right)^2,\e^2\right)\,\,\,\,\, \mbox{as}\,\,\,\,\bar{\kappa}\C^{-1} \to \infty,
\end{gather}
with all other parameters assumed $\O(1)$. When $\bar{\kappa}\C^{-1}$ is finite the tumbling motion is joined by a small relative tangential material oscillation. This periodic phase-lag of the material becomes more pronounced as $\bar{\kappa}\C^{-1}$ is reduced closer to unity, and vanishes as $\bar{\kappa}\C^{-1} \to \infty$, leading ultimately to pure (rigid-body) tumbling motion.

For a given material property contrast, we have now seen that decreasing the shear-rate below a critical value produces, perhaps counter-intuitively, a tumbling motion, while increasing it above this value invites the vesicle to swing. With a very slow background flow the vesicle elongates in the directions of its softest components (or higher spontaneous curvature regions) in a quasi-steady manner - the material is nearly matched to the shape as it rotates like a rigid body. But in a flow with a large shear-rate, material is driven around the surface with larger viscous stresses relative to the elastic stresses, and the stiffer material may be driven past the high curvature regions. High curvature regions can rapidly align with the softer regions via a rapid swing.

Similar transitions from swinging to tumbling have been observed in red blood cells~\citep{ng09}, capsules~\citep{kfs08,Barthes-Biesel91,Barthes-Biesel16} and vesicles even with uniform bending rigidity~\citep{ks06,Lebedev2007,dkss09,dks09} but at smaller reduced areas. In addition, the variation in spontaneous curvature along the surface of a red blood cell has previously been modeled through a simple energy barrier - there too the contrast in material properties revealed a transition from tumbling to swinging \citep{ss07}.

\subsubsection{Matched asymptotic analysis}
Between these two extremes lies a critical value of $\bar{\kappa}\C^{-1}$ which signals a bifurcation from swinging to tumbling. Figure~\ref{fig:swinging_to_tumbling}(b) shows the minimum inclination angle achieved during the periodic dynamics for a range of $\e$ and $\bar{\kappa}\C^{-1}$ found using the full numerical simulations. The bending stiffness variation needed to set off a tumbling dynamics is near unity as $\e \to 0$, (consistent with the much simpler numerical solutions of Eqn.~\eqref{po2}), and is a decreasing function of $\e$. Note that \tcb{$\bkc=\e\bar{\kappa}\Ca^{-1}$ depends on $\e$ in Fig.~\ref{fig:swinging_to_tumbling}b.} A thin band near this bifurcation ridge shows an unexpected result: the membrane's inclination angle can decrease to values less than zero even during a (rather dramatic) swinging motion. Common intuition from single-component membrane dynamics suggests that once the elongated axis has dipped below the x-axis the membrane will surely tumble; this intuition is thus not always correct.

We are therefore led to investigate the regime where $\bar{\kappa}\C^{-1}=1+O(\e)$ and we define $\alpha=(\bkc-1)/\e$ with $\alpha=O(1)$ as $\e \to 0$. For values of $\tau$ where $\po_\tau=O(1)$ as $\e \to 0$, the inclination angle is drifting slowly and an outer solution is derived assuming a regular expansion in $\e$, $\po=\po_{outer}+O(\e)$, resulting in $\po_{outer}(\tau) = 3\pi/8-\tau/4+O(\e)$. The initial value of $\po$ does not appear in the outer solution because the distribution of bending stiffness begins entirely in the $\cos(2\theta)$ mode at $\tau=0$ from Eqn.~\eqref{eqn:prescribedBending}; there is a rapid correction on a timescale $O(\e)$ (just visible near $\tau=0$ in Fig.~\ref{fig:swinging_to_tumbling}(a)) before the outer solution becomes dominant, and memory of the initial state is almost immediately lost.

An inner region of rapid variation in $\po$ emerges when $\po\approx 0$, or when $\tau\approx 3\pi/2$. The scaling of the inner region in $\tau$ and the solution there are found by appealing to dominant balance as $\e \to 0$ \citep{bo13}, leading to the definition of an inner variable $\sigma = (\tau-3\pi/2)/\e^{1/2}$, so that Eqn.~\eqref{po2} reads as
\begin{gather}\label{eq: pos}
\e^{1/2}\po_\sigma = \beta\left(\cos\left(2\po\right)-(1+\e \alpha)\cos\left(2\po+\e^{1/2}\sigma\right)\right),
\end{gather}
\tcb{where $\beta=(1+\lambda)^{-1}Q^{-1}$,} and an Ansatz $\po_{inner}=p_{inner}^{(0)}(\sigma)+\e^{1/2}p_{inner}^{(1)}(\sigma)+O(\e)$. At leading order we find
\begin{gather}\label{pinnereq0}
\frac{d}{d\sigma}p_{inner}^{(0)} = \beta \sigma\sin(2p_{inner}^{0}),
\end{gather}
which has solution
\begin{gather}\label{pin0}
p_{inner}^{(0)}(\sigma)=\tan^{-1}\left(C_0 e^{\beta \sigma^2}\right),
\end{gather}
with $C_0$ an integration constant. However, in order for this inner solution to merge with the outer solution, or
\begin{gather}
\lim_{\tau \to 3\pi/2^{-}} \po_{outer} = \lim_{\sigma \to -\infty} \po_{inner},
\end{gather}
we must have that $C_0=0$. At the next order Eqn.~\eqref{eq: pos} then produces
\begin{gather}\label{pinnereq1}
\frac{d}{d\sigma}p_{inner}^{(1)} = \beta\left(\alpha-\frac{\sigma ^2}{2}-2 \sigma p_{inner}^{(1)}\right),
\end{gather}
and the solution
\begin{gather}\label{pin1}
p_{inner}^{(1)}(\sigma)=-\frac{\sigma}{4}+\PARAM\left(\mbox{erf}(\beta^{1/2}\sigma)+C_1\right)e^{\beta \sigma^2},
\end{gather}
where $C_1$ is an integration constant and
\begin{gather}\label{PARAM}
\PARAM = \frac{1}{8}\left(\frac{\pi}{ \beta}\right)^{1/2}(1-4\beta \alpha).
\end{gather}
The error function, $\mbox{erf}(\beta^{1/2}\sigma)=(2/\sqrt{\pi})\int_0^{\beta^{1/2}\sigma} e^{-t^2}\,dt$, is an odd function which tends towards $-1$ as $\sigma\to -\infty$ and to $1$ as $\sigma\to \infty$. Again the requirement of matching to the outer solution demands that terms which are unbounded as $\sigma\to -\infty$ vanish, leading to $C_1 = 1$. The inner solution alone then represents a composite approximation for $\tau\in(O(\e),3\pi/2]$,
\begin{gather}\label{po_left}
\po(\tau)\sim\frac{3\pi}{8}-\frac{\tau}{4}+\e^{1/2}\PARAM \left[\mbox{erf}\left(\sqrt{\frac{\beta}{\e}} \left(\tau -\frac{3 \pi }{2}\right)\right)+1\right]e^{\beta(\tau-3\pi/2)^2/\e}.
\end{gather}
This solution becomes unbounded when $\tau$ increases beyond $3\pi/2$, so is incapable of merging with the possible outer solutions to the right of $\tau=3\pi/2$, either $7\pi/8-\tau/4$ if $(1-4\beta \alpha)>0$ (a swing) or $-\pi/8-\tau/4$ if $(1-4\beta \alpha)<0$ (a tumble). Instead we continue the solution by solving Eqn.~\eqref{po2} on $\tau\geq 3\pi/2$ using the initial data from the inner solution above, $\po(3\pi/2)=\e^{1/2}\PARAM +O(\e)$.

The solution at leading order is again that in Eqn.~\eqref{pin0} but this time $C_0\neq 0$. After finding the solution at the next order (included as \S\ref{AppendixF}), however, in order to match both the data at $\tau=3\pi/2$ and to merge with an outer solution it becomes clear that $C_0 = O(\e^{1/2})$, and then the equation for $p_{inner}^{(1)}$ in Eqn.~\eqref{pinnereq1} and its general solution in Eqn.~\eqref{pin1} go unchanged. Removing unbounded terms as $\sigma \to \infty$ selects $C_1=-1$, and matching the data at $\tau=3\pi/2$ selects $C_0=\tan\left(2\e^{1/2}\PARAM\right)$, resulting in the following composite solution for $\tau\in[3\pi/2,3\pi/2+2\pi-O(\e^{1/2}))$:
\begin{multline}\label{po_right}
\po(\tau)\sim \frac{3\pi}{8}-\frac{\tau}{4}+\tan^{-1}\left[\tan\left(2\e^{1/2}\PARAM\right)e^{\beta(\tau-3\pi/2)/\e}\right]\\
+\e^{1/2}\PARAM \left[\mbox{erf}\left(\sqrt{\frac{\beta}{\e}} \left(\tau -\frac{3 \pi }{2}\right)\right)-1\right]e^{\beta(\tau-3\pi/2)^2/\e}.
\end{multline}

\begin{figure}
\centering
\includegraphics[width=.6\textwidth]{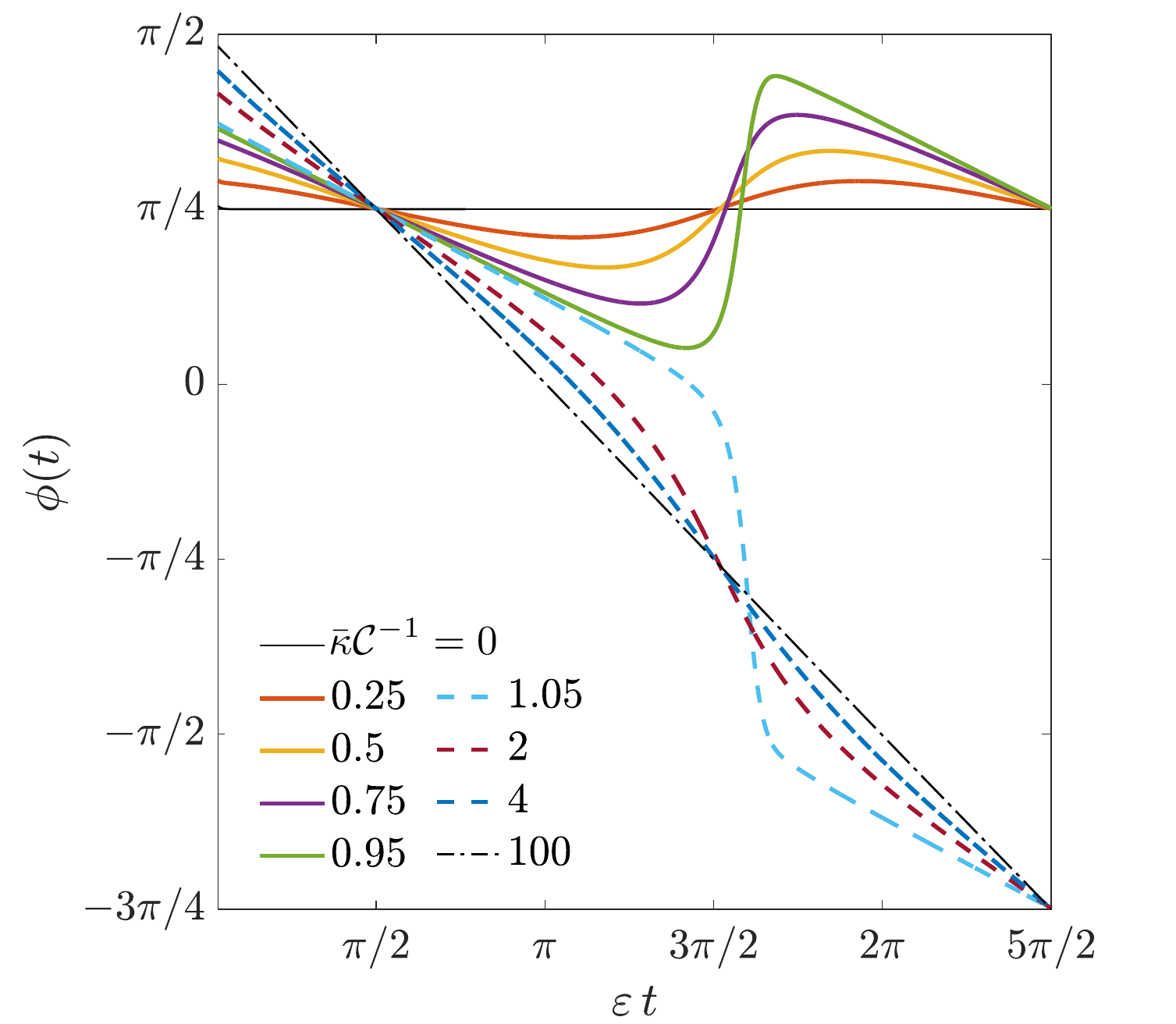}
\caption{(Color online) Theoretical inclination angle (solutions of Eqn.~\eqref{eqn:phitO1M=2}) with spatial period $M=2$ of material property variation, viscosity ratio $\lambda=\mu^-/\mu^+=1$, $\e=10^{-2}$ or $R_A=0.998$, and $Q=3$ for a range of $\bar{\kappa}\C^{-1}$.}
\label{kappa_solutions}
\end{figure}

The critical dependence of the dynamics on the sign of $\bar{\kappa}\C^{-1}-1$ as $\e \to 0$ is thus established, most clearly through the dependence of the argument of $\tan^{-1}$ on the sign of $\PARAM$, and thus on the sign of $(1-4\beta \alpha)$ (and recalling that $\alpha = (\bar{\kappa}\C^{-1}-1)/\e$).

    Since $\beta>0$, if $\bar{\kappa}\C^{-1}<1$ then $(1-4\beta \alpha)>0$ and the solution above shows a rapid return to a positive inclination angle just less than $\pi/2$, representing a dramatic swinging motion. If $\bar{\kappa}\C^{-1}>1$, however, then the dynamics depend on $\beta=[(1+\lambda)Q]^{-1}$. If $\beta>1/(4\alpha)$ then $(1-4\beta \alpha)<0$ and as $\tau$ increases beyond $3\pi/2$ the inclination angle dips rapidly towards negative values and below $-\pi/2$, representing a tumble. If $\beta<1/(4\alpha)$, however, the inclination angle becomes negative as $\tau$ increases away from $3\pi/2$ for a short while, but then for longer times it launches back towards positive values: in this case the membrane's long axis dips below the horizontal, hinting at a tumble, but then rapidly pulls back up into positive inclination angles in a high amplitude swing. Inclination angles from numerical solution of Eqn.~\eqref{eqn:phitO1M=2} with $\e=10^{-2}$ are plotted for a range of $\bar{\kappa}\C^{-1}$ in Fig.~\ref{kappa_solutions}. The approximations in Eqns.~\eqref{po_left}-\eqref{po_right} are visibly indistinguishable (and not shown) from numerical solution of Eqn.~\eqref{eqn:phitO1M=2} in this parameter regime.

The inclination angle equation, Eqn.~\eqref{eqn:phitO1M=2}, only provides a solution for the $O(1)$ behavior of the inclination angle, $\phi^{(0)}(t)$; so while the expressions above are accurate asymptotic solutions to Eqn.~\eqref{eqn:phitO1M=2}, the equation itself is only representing the $O(1)$ behavior of the inclination angle $\phi(t)$. While these analytical representations show remarkable accuracy when compared to the full numerical simulations, seen in Fig.~\ref{fig:swinging_to_tumbling}(a), certain aspects of the full system are delicate. For instance, the analysis above suggests that the critical $\bar{\kappa}\C^{-1}$ beyond which tumbling occurs is an increasing function of $\e$, but this lies in stark contrast to the results of the full numerical simulations shown in Fig.~\ref{fig:swinging_to_tumbling}(b). The analysis above shows, however, that the critical value for the onset of tumbling is indeed $\bar{\kappa}\C^{-1}=1+O(\e)$ as $\e \to 0$, and generally provides accurate dynamics in a very wide variety of settings.

\subsection{The case $M\neq 2$:}

Turning now to the case where $M\neq 2$, the daunting system is rendered harmless upon observation of a periodic steady state in which $P_0(t)$ is constant. According to Eqns.~\eqref{eqn:angen}-\eqref{eqn:bngen} with $P_0$ assumed constant, as $t\to \infty$ we find $a_n=0$ and $b_n=0$ for all $n\notin \{2,M\}$. Meanwhile, as in the constant bending stiffness case, $b_2$ relaxes to an equilibrium value $\rstwoeq$, where $\rstwoeq=\C/\alpha_2=\C \left(3 +\C\,P_0 \right)^{-1}$.

Shape deformations continue periodically in the $M^{\mbox{th}}$ Fourier mode, however, according to Eqns.~\eqref{eqn:angen}-\eqref{eqn:bngen} (upon inserting $c_M(t)$ and $d_M(t)$ from Eqn.~\eqref{eqn:prescribedBending}). At leading order in $\e$ the system is quasi-steady; with $\tau=\e t$ again, we write $\partial_t a_M = \e \partial_\tau a_M$ (similarly for $b_M$). Neglecting a transient relaxation from initial data, to leading order in small $\e$ we find
\begin{gather}
a_M(t)=\frac{-\bar{\kappa}}{\alpha_M}\cos\left(\frac{M\e t}{2}\right),\,\,\,\, b_M(t)=\frac{\bar{\kappa}}{\alpha_M}\sin\left(\frac{M\e t}{2}\right).\label{eqn:ambmnot2}
\end{gather}
Simply, then, in the periodic steady state we have $a_M^2+b_M^2 = \bar{\kappa}^2/\alpha_M^2$, and $a_M c_M+ b_M d_M =-\bar{\kappa}^2/\alpha_M$. As both are constant, along with the constant value of $b_2$ in the limit as $t\to \infty$, upon inspection of $P_0$ in Eqn.~\eqref{eqn:pressure} we verify the consistency of this result: $P_0$ is indeed constant in this periodic steady state. Since $\rstwoeq$ is determined purely by the constraint of constant area, from Eqn.~\eqref{eqn:rstwoeqRA}, the pressure jump $P_0$ associated with these dynamics is the same as that in the constant bending case. Moreover, the deformation parameter and inclination angle in the $M\neq 2$ case are also unchanged. The membrane simply elongates in the direction of the principle axis of the straining flow while shape oscillations in the $M^{th}$ mode traverse along this constant background geometry in a trembling dynamics.

\begin{figure}
\centering
\includegraphics[width=\textwidth]{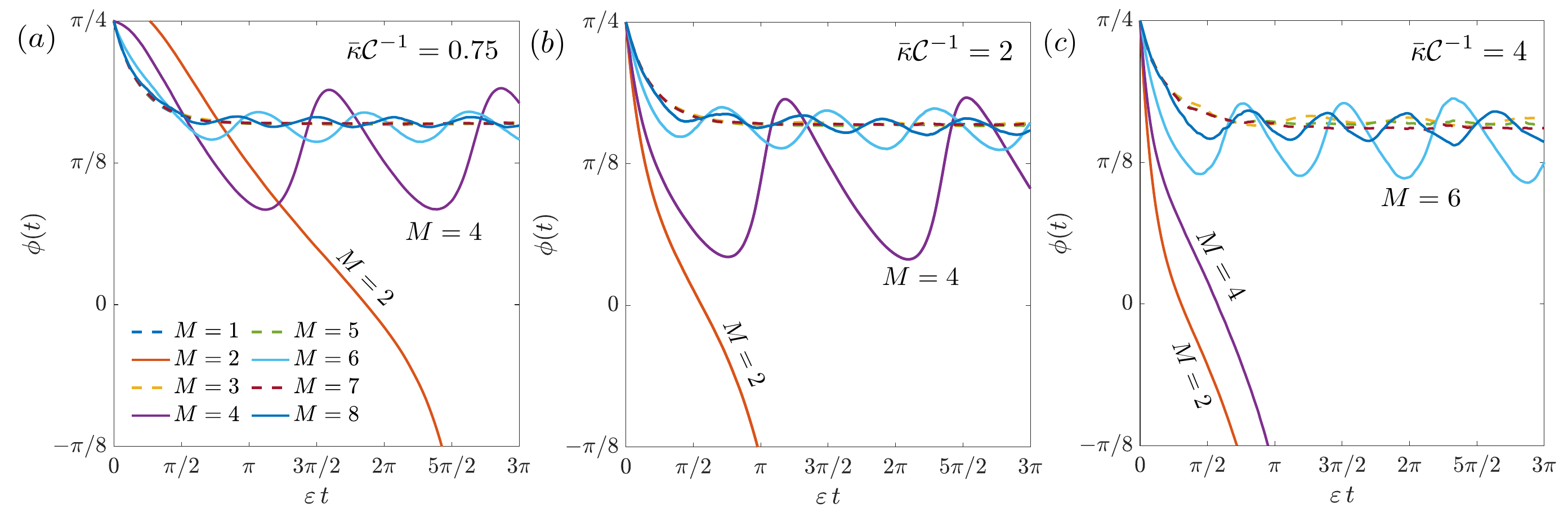}
\caption{(Color online) Inclination angle dynamics with bending stiffness variations in the $M^{th}$ spatial mode for $M\in\{1, 2, ..., 8\}$ with $\e=0.1$ or $R_A=0.865$ and $\lambda=1$ for three variation amplitudes, from the full numerical simulations. (a) $\bar{\kappa}\C^{-1}=0.75$; (b) $\bar{\kappa}\C^{-1}=2$; and (c) $\bar{\kappa}\C^{-1}=4$. Tumbling is observed in small, even modes. (See also Supplemental Movies M4-M6).}
\label{fig: Mplots}
\end{figure}

When $\bar{\kappa}\C^{-1}$ is sufficiently large, interactions between the modes of bending stiffness can no longer be neglected (i.e. our simple specification of $\kappa(\theta)$ in Eqn.~\eqref{eqn:prescribedBending} becomes inaccurate). Full numerical simulations are used to explore this challenging region of parameter space. Figure~\ref{fig: Mplots} shows the inclination angles computed using the full numerical simulations for $M\in\{1, 2, ..., 8\}$, with $\e=0.1$ and $\lambda=1$ fixed, for three different bending stiffness variations, $\bar{\kappa}\C^{-1}=0.75$, $\bar{\kappa}\C^{-1}=2$, and $\bar{\kappa}\C^{-1}=4$. The $M=2$ mode results in tumbling in all three cases, consistent with Fig.~\ref{fig:swinging_to_tumbling}(b). The swinging amplitude with even $M$ values increases with increasing $\bar{\kappa}\C^{-1}$, however, and the $M=4$ case transitions from swinging to tumbling for some $\bar{\kappa}\C^{-1}\in(2,4)$. Supplemental Movies M4-M6 show the dynamics of vesicles with $M\in\{1, 2, ..., 8\}$ represented in Fig.~\ref{fig: Mplots}(a-c).

In the simulated dynamics we observe membrane swinging for small, even $M$, but not odd $M$, or large even $M$ with an insufficiently large value of $\bar{\kappa}\C^{-1}$. When $M$ is even the two regions of largest curvature have a symmetric interaction with the membrane, and elongation in the direction of the softer material reduces the energy at both ends. Bending stiffness information in the $M=4$, mode, for instance, bleeds into the $M=2$ mode, which interacts directly with the extensional part of the background flow and can lead to tumbling, as discussed in the previous section. When $M$ is odd, however, the large curvature regions have an asymmetric interaction with the membrane; reorientation of the elongated axis which would reduce the bending energy on one end would increase it on the other end. Finally, when $M$ is large, either even or odd, averaging results in convergence to the case of constant bending stiffness, and departures from the inclination angle chosen by the principal axis of the background flow, $\pi/4$ as $\e \to 0$, become negligible. It remains to be seen whether a sufficiently large $\bar{\kappa}\C^{-1}$ can result in tumbling for any even $M$; extremely stiff regions do not pass easily across high curvature regions, suggesting that tumbling might ensue for very large values of $\bar{\kappa}\C^{-1}$, but high spatial frequency averaging suggests convergence to pure tank-treading as in \S\ref{S: constant bending}. The answer may well depend on the reduced area and viscosity ratio. We leave this intriguing question for future inquiry.

\section{Discussion}\label{S: conclusion}

The material property variations along the surface of a multicomponent vesicle can impact the vesicle dynamics in a background flow differently depending on the spatial modes of its distribution, the magnitude of those variations, and even the parity of the number of domains. Small amplitude variations in material properties lead to periodic oscillations of a pure tank-treading steady state about an inclination angle of $\pi/4$; large variations can result in a rigid body tumbling mode with a constant rotation rate; and an intermediate regime shows a bifurcation from swinging with tank-treading to tumbling with periodic material phase-lag. As the membrane becomes more deflated the critical value of $\bar{\kappa}\C^{-1}$ required for the vesicle to tumble is found to decrease, with $\bar{\kappa}\C^{-1}$ approaching 1 as $\e$ approaches 0. As a general principle, the vesicle has a tendency to elongate so that the softer parts of the membrane sit in the regions of largest curvature, while the background flow tends to elongate the vesicle along the principal axis with a fixed inclination angle of $\pi/4$. When these two directions are not aligned, swinging, or even tumbling, ensues. That the capillary number is highly sensitive to the vesicle size may be of use to experimental realizations of the results described in this paper.

Although we have focused on variations in bending stiffness, at leading order we find the same shapes, dynamics, and bifurcation from swinging to tumbling when considering variations in spontaneous curvature instead. Eqns.~\eqref{eqn:angen}-\eqref{eqn:bngen} indicate that when the preferred mean curvature along the membrane, $\tilde{H}_0$, is not unity, the effects of bending stiffness variations and spontaneous curvature variations are indistinguishable for each Fourier mode. A model linking the two (e.g. if bending stiffness is proportional to spontaneous curvature for a given lipid species) may then be necessary to make claims about material properties in full using this passive means of probing membrane composition. If $\tilde{H}_0=1$, however, then only spontaneous curvature variations enter at first order in $\e$. 

Replacing bending stiffness by spontaneous curvature, if $\tilde{H}_0 =0$ the transition between tumbling and swinging for the $M=2$ spatial mode is predicted at the critical value $a[\tilde{H}]\C^{-1}=1$ as $\e\to 0$, with $[\tilde{H}]$ the curvature contrast. Estimating the spontaneous curvature variations of a red blood cell to be roughly $[\tilde{H}]=0.5\mu$m$^{-1}$ with $a\approx 3\mu$m, and with $\Ca\approx \g/4$ from \S\ref{sec: nondim}, the theoretical prediction is that the bifurcation should appear near $\g \approx 2$s$^{-1}$. This is very close to the shear-rates used in experiments showing the onset of this transition \citep{afv07,av08}.

\tcb{In the fully three-dimensional system, material domains are not confined to motion in the flow direction only and this may result in a substantial departure from the results described herein in certain regimes. Particularly when slow motions yield to sudden reorganization, as in a rapid swing or tumbling event, the addition of such an escape direction may prove critical. But some material properties cannot so easily be disturbed, for instance the spontaneous curvature of a red blood cell provided by the scaffolding of its spectrin network \citep{dls03,hm15}. That the transition from tumbling to swinging in red blood cells appears to be predicted already using this two-dimensional analysis, however, is intriguing.}

The distribution of membrane domains is of substantial biological importance. Membrane heterogeneity can impact fundamental cellular functions such as signal transduction and membrane trafficking~(\cite{Edidin03,st00, Maxfield02}), and improper composition can cause diseases such as Alzihmers~(\cite{vt10,ra12}). The predictions of this work suggests a means of determining not only the constant material properties of a membrane or vesicle using a background flow, which has been an experimentally viable method for decades, but now also of determining material property variations by linking time-series dynamics to spatial material variations, and even the possibility of using a simple pressure probe near such a swinging, tumbling, and trembling membrane. With good fortune, these predictions will be of use for measuring heterogeneous membrane properties using only viscous stresses in the near future.

Declaration of Interests: The authors report no conflict of interest.

\acknowledgements
S.E.S. acknowledges the support of the NSF/NIH (DMS-1661900, DMR-2003819).

\appendix

\section{Stream-function and incompressibility in polar coordinates}\label{AppendixA}
The (dimensional) biharmonic equation in polar coordinates $(r,\theta)$ has a general solution known as the Michell solution \citep{Michell99}. Neglecting terms which are non-periodic in $\theta$, the biharmonic equations inside ($-$) and outside ($+$) the vesicle are solved by
\begin{multline}
\psi^\pm = A_0^\pm r^2+B_0^\pm r^2\ln(r)+C^\pm_0\ln(r)+\left(A^\pm_1 r+B^\pm_1 r^{-1}+C^\pm_1 r^3+D^\pm_1 r\ln(r)\right)\cos(\theta)\\
+\left(E^\pm_1 r+F^\pm_1 r^{-1}+G^\pm_1 r^3+H^\pm_1 r\ln(r)\right)\sin(\theta)\\
+\sum_{n=2}^\infty \left(A^\pm_n r^n+B^\pm_n r^{-n}+C^\pm_n r^{n+2}+D^\pm_n r^{2-n}\right)\cos(n \theta)\\
+\sum_{n=2}^\infty\left(E^\pm_n r^n+F^\pm_n r^{-n}+G^\pm_n r^{n+2}+H^\pm_n r^{2-n}\right)\sin(n \theta).\label{Eqn:gensol}
\end{multline}
The coefficients above are determined instantaneously in time by demanding that $\psi^{-}$ and its derivatives are bounded at the origin, convergence to the far-field limit ($\psi^+ \to \g r^2\sin^2(\theta)/2$ as $r \to \infty$), continuity of velocity across the membrane boundary, $[\nabla \psi]_S=\bm{0}$, traction balance (see \S\ref{AppendixB}), and surface inextensibility along the membrane, $\nabla_s \cdot \u|_S=0$, where $ \nabla_s$ is the surface del operator,
\begin{gather}
 \nabla_s  = \shat\left(\shat\cdot\nabla \right)  =\left(\that+\e \rho_\theta\rhat+O(\e^2) \right)\left(\that+\e \rho_\theta\rhat   +O(\e^2) \right)\cdot \left(\rhat \partial_r +\that\frac{1}{r}\partial_\theta \right)\nonumber\\
=\that\frac{1}{r} \partial_\theta +\e  \left(\that \rho_\theta \partial_r +\rhat \frac{\rho_\theta}{r} \partial_\theta  \right)+\O(\e^2).\label{eq:sdiv}
\end{gather}
Inextensibility is given in terms of the radial and azimuthal velocity components $u_r$ and $u_\theta$ by
\begin{gather}
\nabla_s\cdot \u\Big|_{S} = \nabla_s\cdot\left(u_r \rhat + u_\theta \that\right)\Big|_{S} =\frac{1}{r}\left(\partial_\theta u_{\theta}  +u_r\right)\Big|_{r=1}+\O(\e|\u|) =0\label{eq:sdivuOeps2}.
\end{gather}
In terms of the stream-function, the relations $u_r = \psi_\theta/r$ and $u_\theta=-\psi_r$ are inserted into the above,
\begin{gather}
\nabla_s\cdot \u\Big|_{S}=\frac{1}{r}\left(-\partial_{ r\theta}\psi+\frac{1}{r} \partial_\theta\psi\right) \Big|_{r=1}+\O(\e \psi)=0.
\end{gather}
More terms above are kept to extend the approximation to higher order.

\section{Traction balance asymptotics}\label{AppendixB}
Traction balance is demanded order by order in the small parameter $\e$. Regular perturbation expansions for the (dimensionless) stream-function, $\psi=\psi^{(1)}+\e \psi^{(2)}+...$, pressure $p=p^{(0)}+\e p^{(1)}+\e^2 p^{(2)}+...$ and viscous traction $\f =\f^{(0)}+ \e \f^{(1)}+...$ are assumed. The dimensionless viscous traction is given at leading order (from Eqn.~\eqref{eq: traction}) by $\f^{(0)}=-[p^{(0)}]\nhat$, and the contribution at first order in $\e$ is given by
\begin{multline}
\f^{(1)} = -\left([p^{(1)}+\rho(\theta,t) \partial_r p^{(0)}]+2 [\partial_\theta\psi^{(1)}- \partial_{r\theta}\psi^{(1)}]^{(\lambda)} \right) \nhat \\
+[ \partial_{\theta\theta}\psi^{(1)}+\partial_{r}\psi^{(1)}-\partial_{rr}\psi^{(1)} ]^{(\lambda)}  \shat,
\end{multline}
where we have defined a jump operator which incorporates the viscosity ratio,
\begin{gather}
[\psi]^{(\lambda)} =  \psi^{+}-\lambda\psi^{-}\Big|_S.
\end{gather}
This viscous traction must balance with the elastic traction. At leading order, traction balance in the tangential and normal directions returns
\begin{gather}
\partial_\theta T^{(0)} =0,\\
-T^{(0)}-[p^{(0)}]=0.
\end{gather}
Hence $T^{(0)}=-[p^{(0)}] =:P_0(t)$, the leading order isotropic tension is balanced with the leading order pressure jump across the interface, a dimensionless statement of an elastic Young-Laplace law. At the next order in $\e$, traction balance in the tangential and normal directions are given by
\begin{gather}
\partial_\theta T^{(1)}-\C^{-1} \left((\tilde{H}_0  -1)\partial_{\theta}\kappa+\partial_{\theta}\zeta+\partial_{\theta}^3\rho+\partial_{\theta}\rho\right)+[\partial_{\theta\theta}\psi^{(1)}+\partial_{r}\psi^{(1)}-\partial_{rr}\psi^{(1)}]^{(\lambda)} =0,\label{Eqn:tractionBalanceTangential}
\end{gather}
with $\C = \Ca/\e$, $\kappa$ and $\zeta$ the first-order material property variations defined in Eqn.~\eqref{eq: achi}, and
\begin{multline}
-T^{(1)}+\left(\rho+\partial^2_{\theta}\rho\right)P_0-\C^{-1} \left((\tilde{H}\tilde{H}_0  -1)\partial^2_{\theta} \kappa+\partial^2_{\theta}\zeta+\partial^4_{\theta}\rho+\partial^2_{\theta}\rho\right)\\
-\left([p^{(1)}+\rho(\theta,t) \partial_r p^{(0)}]+2 [\partial_\theta\psi^{(1)}- \partial_{r\theta}\psi^{(1)}]^{(\lambda)} \right)=0.\label{Eqn:tractionBalanceNormal}
\end{multline}
In the limit of infinite bending capillary number (i.e. zero bending stiffness) these expressions are consistent with those provided by \cite{zrs87}. The membrane length and area constraints, enforced out to second-order in $\e$ as $\e \to 0$, are used to determine pressure jump at the interface $P_0$ at leading order (or the isotropic tension $T^{(0)}$), leading to the expression in Eqn.~\eqref{eqn:pressure}.

\section{First-order solution}\label{subsection: solving stream}\label{AppendixC}

Equations~\eqref{Eq:biharmonicStream}, \eqref{Eqn:continuity}, \eqref{Eqn: inextensibility}, \eqref{Eqn:tractionBalanceTangential} and \eqref{Eqn:tractionBalanceNormal} are solved simultaneously for the dimensionless first-order stream-function $\psi^{(1)}$ (via Eqn.~\eqref{Eqn:gensol}, properly scaled) and pressure $p^{(1)}$ both inside and outside the vesicle, and for the first-order membrane tension, $T^{(1)}$. The resulting stream-functions are given by
\begin{multline}
\psi^{(1)-}=\frac{r^2}{4}-\frac{r^2(3-r^2)}{4(1+\lambda)}\cos(2\theta)\\
+\sum_{n=2}^{\infty} r^n\left((n+1)-(n-1)r^2\right)\left(B_n \cos(n\theta)
-A_n\sin(n\theta)\right)\label{eqn:psim}
\end{multline}
and
\begin{align}
\psi^{(1)+}&=\frac{r^2}{4}-\left(\frac{\lambda+r^2(1-2\lambda)+r^4(1+\lambda)  }{4r^2(1+\lambda) } \right)\cos(2\theta)\nonumber \\
&+\sum_{n=2}^{\infty}\left((n+1)r^{2-n}-(n-1)r^{-n}\right)\left(B_n\cos(n\theta)-A_n\sin(n\theta)\right),\label{eqn:psip}
\end{align}
(note that $n=2$ terms are also in the summation), where $A_n$, $B_n$ are given in \eqref{eqn:JnKn}. With $p^{(0)+}=p_\infty$ and $p^{(0)-}=p_\infty+P_0$ the (spatially constant) leading-order pressure fields, with $P_0$ given in Eqn.~\eqref{eqn:pressure}, the first-order pressure fields are
\begin{gather}
p^{(1)-}=\Pi-\frac{3\lambda r^2}{1+\lambda}\sin(2\theta) + 4\lambda \sum^{\infty}_{n=2}(n^2-1)r^n  \left( B_n \sin(n\theta)+ A_n \cos(n\theta)\right)
\end{gather}
and
\begin{gather}
p^{(1)+}=\frac{(2\lambda-1)}{r^2(1+\lambda)}\sin(2\theta) +4 \sum^{\infty}_{n=2}r^{-n}(n^2-1)\left(B_n \sin(n\theta) -A_n \cos(n\theta)\right),
\end{gather}
where $\Pi$ is a constant. Finally, the membrane tension at first-order is given by
\begin{gather}
T^{(1)}=\Pi -\sin(2 \theta)-\sum^{\infty}_{n=1} \left(4(1+\lambda)B_n-P_0b_n\right)\sin(n\theta)+ \left(4(1+\lambda)A_n-P_0a_n\right)\cos(n\theta).
\end{gather}
The free constant $\Pi$ appears in both $p^{(1)-}$ and $T^{(1)}$, indicating an ambiguity which is understood upon interpretation of the pressure and tension fields as Lagrange multipliers which enforce fluid and membrane incompressibility and inextensibility, respectively, and recalling that the two are linked by the Young-Laplace law. The value of $\Pi$ has no bearing on the dynamics.

\section{From the stream-function to the surface velocity}\label{AppendixD}
For a given station in arc-length $s$, the no-slip condition is written as $\partial_t \r(s,t) = \u(\r(s,t),t)$; to focus on fixed values of $\theta$ we write $\r(s,t)=\r(s(\theta,t),t)=r(\theta,t)\rhat(\theta)$. Then noting that
\begin{gather}
\frac{\partial \r}{\partial t}\Big|_{s} = \frac{d \r}{dt} - \frac{\partial \r}{\partial s}\frac{\partial s}{\partial t}\Big|_{\theta}= \frac{d \r}{dt} - \shat \frac{\partial s}{\partial t}\Big|_{\theta},
\end{gather}
dotting with the normal vector removes the need to determine $\partial_t s$ for fixed $\theta$:
\begin{gather}
\nhat \cdot \frac{\partial \r}{\partial t}=\nhat \cdot \frac{d \r}{dt} =\nhat \cdot \u\Big|_S,
\end{gather}
and thus $\partial_t \left(r(\theta,t) \rhat\right)\cdot \nhat=  \nhat \cdot \u|_S$. Then with $\u = u_n \nhat + u_s \shat =u_r \rhat + u_\theta \that$, and all dimensionless velocities expanded as $\u = \u^{(1)}+\e \u^{(2)}+...$,
\begin{gather}
\frac{d}{dt}\left(r(\theta,t) \rhat\right)\cdot \nhat = \left(\e \rho_t + \e^2 \rhotwo_t+\O(\e^3)\right) \left(1+O(\e^2)\right) =  \e \rho_t + \e^2 \rhotwo_t+\O(\e^3),
\end{gather}
and
\begin{gather}
\nhat \cdot \u\Big|_S = \left(\rhat  - \e \rho_\theta \that+O(\e^2)\right)\cdot \u \Big|_S = \left(u_r^{(1)} +\e\left(u_r^{(2)}- \rho_\theta u_\theta^{(1)}\right)+\O(\e^2)\right)\Big|_S.
\end{gather}
Recall that the dimensional velocity is scaled by $\g$, so that a dimensionless velocity $\u^{(1)}$ which is $O(1)$ as $\e \to 0$ corresponds to a dimensional velocity $\g \u^{(1)}$ which is $O(\e)$ as $\e \to 0$. Since the velocities in the radial and azimuthal directions are given by
\begin{gather}
u_r \Big|_S= \frac{1}{r}\psi_\theta(r,\theta) \Big|_{r=1+\e\rho+\O(\e^2)}= \psi_\theta^{(1)}\Big|_{r=1}+\e\left(\psi_\theta^{(2)}+\rho\left( \psi_{r\theta}^{(1)}- \psi_\theta^{(1)}\right)\right)\Big|_{r=1}+\O(\e^2),\\
u_\theta \Big|_S= -\psi_r(r,\theta) \Big|_{r=1+\e\rho+\O(\e^2)}= -\psi_r^{(1)}\Big|_{r=1}-\e\left(\psi_r^{(2)}+\rho \psi_{rr}^{(1)}\right)\Big|_{r=1}+\O(\e^2),
\end{gather}
the velocity in the surface normal direction may be written as
\begin{gather}
\nhat \cdot \u\Big|_{S} = \psi_\theta^{(1)}\Big|_{r=1}+ \e\left(\psi_\theta^{(2)}+\rho( \psi_{r\theta}^{(1)}- \psi_\theta^{(1)})+\rho_\theta \psi_r^{(1)}\right)\Big|_{r=1}+\O(\e^2).
\end{gather}
Hence, since the dimensional time is scaled by $\e/\g$,
\begin{gather}
\rho_t = u_n^{(1)}\Big|_S = u_r^{(1)}\Big|_S =  \psi^{(1)}_\theta\Big|_{r=1},\,\,\, \mbox{and}\\
\rhotwo_t = u_n^{(2)}\Big|_S = u_r^{(2)}-\rho_\theta u_\theta^{(1)}\Big|_S= \psi_\theta^{(2)}+\rho\left( \psi_{r\theta}^{(1)}- \psi_\theta^{(1)}\right)+\rho_\theta \psi_r^{(1)}\Big|_{r=1}.
\end{gather}
Since the gradient of the stream-function is continuous across the membrane boundary, either $\psi^{+}$ or $\psi^{-}$ may be inserted into the above without ambiguity. Using the results of \S\ref{AppendixC}, the normal and tangential components of the velocity are then given by Eqns.~\eqref{eqn:un}-\eqref{eqn:us}.

\section{Inertia tensor, deformation parameter, and inclination angle}\label{AppendixE}

The deformation parameter, $D=(L_1-L_2)/(L_1+L_2)$, is defined using the axis lengths $2L_1$ and $2L_2$ of the ellipse which shares the same inertia tensor. With $\Omega$ denoting the vesicle's interior, the inertia tensor is defined as
\begin{gather}\label{eq:InertiaTensor}
\mathbb{I} = \int_{\Omega}\left(\begin{array}{cc} y^2 & -x y \\ -x y & x^2\end{array}\right)\,dx\,dy.
\end{gather}
When $\Omega$ is the interior of an ellipse with major and minor axis lengths $2L_1$ and $2L_2$, respectively, oriented with its major axis at an angle $\theta$ relative to the x-axis, this tensor has eigenvalues $\lambda_1 = \pi L_1^3 L_2/4$ and $\lambda_2 = \pi L_1 L_2^3/4$, with associated eigenvectors $\bm{v}_1 = (\sin^2\theta,-\sin(2\theta)/2)$ and $\bm{v}_2 = (\cos^2\theta,\sin(2\theta)/2)$. In terms of the eigenvalues of the inertia tensor, then, $L_1 = (4/\pi)^{1/4}(\lambda_1^3/\lambda_2)^{1/8}$ and $L_2 = (4/\pi)^{1/4}(\lambda_2^3/\lambda_1)^{1/8}$, the deformation parameter is given by
\begin{gather}\label{eq:ddef}
D = \frac{L_1-L_2}{L_1+L_2} = \frac{\lambda_1^{1/2}-\lambda_2^{1/2}}{\lambda_1^{1/2}+\lambda_2^{1/2}},
\end{gather}
and the inclination angle may be recovered from $\bm{v}_2$ via $\theta = \tan^{-1}(\yhat\cdot \bm{v}_2/\xhat\cdot \bm{v}_2)$.

Considering the general membrane boundary $S$, parameterized as in \S\ref{S: model}, the inertia tensor above instead has eigenvalues
\begin{gather}
\lambda_1 =\frac{1}{4} \left\{\pi+2 \pi  \e \sqrt{a_2^2+b_2^2}+\e ^2 \left(3 \left(a_2^2+b_2^2\right)+2\frac{a_2^{(2)}a_2 +b_2^{(2)}b_2 }{\sqrt{a_2^2+b_2^2}}\right)\right\}+O(\e^3),\\
\lambda_2 =\frac{1}{4} \left\{\pi-2 \pi  \e \sqrt{a_2^2+b_2^2}+\e ^2 \left(3 \left(a_2^2+b_2^2\right)-2\frac{a_2^{(2)}a_2 +b_2^{(2)}b_2 }{\sqrt{a_2^2+b_2^2}}\right)\right\}+O(\e^3),
\end{gather}
and then Eqn.~\eqref{eq:ddef} produces the deformation parameter in Eqn.~\eqref{eqn:TaylorDeform}. The eigenvector associated with $\lambda_2$ has components
\begin{gather}
\xhat\cdot\bm{v}_2 = b_2^3+a_2 b_2 \left(\sqrt{a_2^2+b_2^2}+a_2\right)-\e  \left(\sqrt{a_2^2+b_2^2}+a_2\right) (b_2^{(2)} a_2-a_2^{(2)}b_2),\\
\yhat\cdot \bm{v}_2=b_2^2 \sqrt{a_2^2+b_2^2},
\end{gather}
and then $\tan^{-1}(\yhat\cdot \bm{v}_2/\xhat\cdot \bm{v}_2)$ returns the inclination angle in Eqn.~\eqref{eqn:incAngle}.

\section{General solution to the inner expansion equations}\label{AppendixF}
The general solution to Eqn.~\eqref{pinnereq0} is
\begin{gather}
p_{inner}^{(0)}(\sigma)=m\pi+\tan^{-1}\left(C_0 e^{\beta \sigma^2}\right).
\end{gather}
for $m$ an integer and $C_0$ an integration constant. At the next order Eqn.~\eqref{eq: pos} then produces
\begin{gather}
\frac{d}{d\sigma}p_{inner}^{(1)} = \beta\left(\frac{1-C_0^2 e^{2 \beta  \sigma ^2}}{1+C_0^2 e^{2 \beta  \sigma ^2}}\right)\left(\alpha-\frac{\sigma ^2}{2}-2 \sigma p_{inner}^{(1)}\right),
\end{gather}
and the solution
\begin{multline}
p_{inner}^{(1)}(\sigma)=\frac{C_1 e^{\beta  \sigma ^2}}{1+C_0^2e^{2 \beta  \sigma ^2}}-\frac{\sigma}{4}\\
+\frac{\pi ^{1/2} e^{\beta  \sigma ^2}}{8 \beta^{1/2}\left(1+C_0^2e^{2 \beta  \sigma ^2}\right)}\left(C_0^2 (1+4 \alpha\beta)\mbox{erfi}\left(\beta^{1/2}\sigma \right)+(1-4 \alpha \beta) \mbox{erf}\left(\beta^{1/2} \sigma \right)\right),
\end{multline}
where $C_1$ is an integration constant. The imaginary error function, $\mbox{erfi}(\beta^{1/2}\sigma)=(2/\sqrt{\pi})\int_0^{\beta^{1/2}\sigma} e^{t^2}\,dt$, tends towards $e^{\beta \sigma^2}/\sqrt{\pi\beta \sigma^2}$ as $|\sigma|\to \infty$.

The approach in \S\ref{subsec: bif} requires that $p_{inner}^{(0)}(0)+\e^{1/2} p_{inner}^{(1)}(0)=\e^{1/2}\PARAM+O(\e)$, with $\PARAM$ defined in \eqref{PARAM}, resulting in
\begin{gather}
m\pi+ \tan^{-1}\left(C_0\right)+\frac{\e^{1/2}C_1}{1+C_0^2} = \e^{1/2}\PARAM+O(\e).
\end{gather}
Here we see that $C_0$ cannot be $O(1)$ as $\e\to 0$, as in this case matching the initial data at $\tau=3\pi/2$ is not possible. But $C_0$ cannot be zero or else either matching to the data above, or merging with the outer solution as $\sigma\to \infty$, is not possible. The equation above is then to be seen as a signal that $C_0$ is in fact $O(\e^{1/2})$ as $\e \to 0$.

\bibliographystyle{jfm}
\bibliography{mvpt}

\end{document}